%% file: main.tex
\documentclass[review,12pt,sort&compress]{elsarticle}

\usepackage{epstopdf}                     

\usepackage{amssymb}
\usepackage{color}
\usepackage[left]{lineno}

\journal{Powder Technology}

\begin{document}

\begin{frontmatter}



\title{Mimicking layer inversion in solid-liquid fluidized beds in narrow tubes \tnoteref{label_note_copyright} \tnoteref{label_note_doi}}

\tnotetext[label_note_copyright]{\copyright 2019. This manuscript version is made available under the CC-BY-NC-ND 4.0 license http://creativecommons.org/licenses/by-nc-nd/4.0/}

\tnotetext[label_note_doi]{Accepted Manuscript for Powder Technology, v. 364, p. 994-1008, 2020, DOI:10.1016/j.powtec.2019.09.089}


\author[label_david]{Fernando David C\'u\~nez Benalc\'azar}
\author[label_erick]{Erick de Moraes Franklin}

\address{School of Mechanical Engineering, University of Campinas - UNICAMP\\
Rua Mendeleyev, 200, Campinas - SP, CEP: 13083-860\\
Brazil}

\address[label_david]{e-mail: fernandodcb@fem.unicamp.br}
\address[label_erick]{phone: +55 19 35213375\\
e-mail: franklin@fem.unicamp.br, Corresponding Author}

\begin{abstract}
\begin{sloppypar}
This paper investigates experimentally and numerically the dynamics of solid particles during the layer inversion of binary solid-liquid fluidized beds in narrow tubes.  Layer inversion can happen in solid classifiers and biological reactors, where different solid particles coexist and segregation by diameter and density occurs. The fluidized beds were formed in a 25.4 mm-ID pipe and consisted of alumina and aluminum beads with diameters of 6 and 4.8 mm, respectively. We placed initially the lighter particles on the bottom in order to force an inversion of layers, mimicking the layer inversion mechanism. In the experiments, we filmed the bed with a high-speed camera and tracked individual beads along images, while in numerical simulations we computed the bed evolution with a CFD-DEM (computational fluid dynamics - discrete element method) code. We found the distances traveled by individual particles during the inversion and the characteristic time for layer inversion.
\end{sloppypar}
\end{abstract}


\end{frontmatter}


\include{text}






\bibliographystyle{elsarticle-num}
\bibliography{references1}







\end{document}

%% file: text.tex
\section{Introduction}
\label{intro}

Solid-liquid fluidized beds (SLFB) consist basically in a suspension of solid particles in an upward liquid flow, being frequently used in petrochemical, food and biochemical industries for their high rates of mass and heat transfers between the liquid and solids. In many practical applications, such as solid classifiers and biological reactors, different solid particles coexist and segregation by diameter and density occurs. The segregation leads to granular layers composed of single solid species, which may or not be desirable \cite{Epstein_1, Epstein_3}. Depending on the grains within the bed, layer inversions of segregated solids take place \cite{Hancock, Epstein_2, Epstein_3}.

Layer inversion may happen in binary beds where the smaller particles have higher density than the larger ones \cite{Epstein_2, Epstein_3}. For these beds, by increasing the liquid velocity slightly above the minimum fluidization, two separated layers form where the smaller particles are on the bottom and the larger on the top of the bed. If the liquid velocity continues to increase further, at some point the layers invert, the smaller particles migrating to the top and the larger ones to the bottom. The first models for layer inversion were based on integral balances over the entire bed, where all the solid-liquid and solid-solid interactions were embedded in the model through the use of closure correlations. Most of these models described the bulk density of each species as a function of the superficial velocity of the liquid, and then searched for an intersection of the functions of each species. Epstein and Leclair \cite{Epstein_2} proposed one of the first quantitative expressions for the layer inversion in binary SLFB. Based on momentum and mass conservations for each species, together with experimental correlations, the authors proposed a model to predict the necessary and sufficient conditions for layer inversion, and, in addition, its critical fluid velocity. Epstein and Leclair \cite{Epstein_2} compared the model results with the experiments of Epstein et al. \cite{Epstein_1}, and the agreement between them was good.

Gibilaro \cite{Gibilaro} proposed modifications in previous layer inversion models for binary SLFB, that were based on bulk densities of single species, in order to account for the presence of a mixture of species in the bottom layer, as observed in the experiments of Moritomi et al. \cite{Moritomi}. The model results compared well with their experiments and that presented by Moritomi et al. \cite{Moritomi} and Epstein and Leclair \cite{Epstein_2}. Chun et al. \cite{Chun} investigated experimentally layer inversions in binary SLFB in the presence of gas flows. The authors used a 1.2 m-long, 210 mm-ID semi-cylindrical duct, where they formed SLFB with and without the presence of gas flow. The beds consisted of polymer beads with $d$ = 3.2 mm and $\rho_p$ = 1280 kg/m$^3$  and glass beads with $d$ = 0.385 mm and $\rho_p$ = 2500 kg/m$^3$, where $d$ is the grain diameter and $\rho_p$ is its density, and the fluids employed were water and air. Chun et al. \cite{Chun} found that the injection of gas flows in a SLFB undergoing layer inversion decreases the threshold velocity for the inversion. The authors proposed that this reduction in the critical velocity is due to an increase in the interstitial velocity of the liquid, which in its turn is caused by a decrease in the liquid cross-sectional area by the presence of gas.  

More recently, Di Maio and Di Renzo \cite{Dimaio} proposed a model based on integral balances for layer inversion in binary SLFB. The authors considered the drag force for a polydisperse granular system, and the balance was solved for the void fraction at inversion. Their model returns the inversion voidage by using as input the diameter, density and volume fraction of the two species. Once the inversion voidage is obtained, the inversion velocity can be obtained by using expressions relating the fluid velocity to the void fraction at inversion. They compared the model results with experiments and the agreement between them was good.

Since the last decade, grain-resolved numerical simulations are being conducted to investigate layer inversion in SLFB. Detailed information that are relatively easy to obtain from numerical simulations, such as the local fluid and particle velocities, local volume fractions, and intermixing levels, are sometimes difficult to be measured in experiments. This kind of information is important to understand mass, momentum and heat transfers in fluid-solid flows, making of numerical simulations an interesting source of data. Eulerian two-phase methods were among the first approaches employed in three-dimensional simulations of fluidized beds. While those methods give reasonable results for problems where the number of particles is relatively large \cite{Cornelissen, wang}, they may fail for the narrow bed case for which confinement effects are strong. Comparisons even for large beds show that Euler-Lagrange methods give similar \cite{Patankar, Ghatage, Alobaid} or better \cite{Chiesa, Vegendla, Adamczyk} results than Euler-Euler methods. For SLFB in narrow tubes, the discrete nature of solids must be accounted for; therefore, methods based on DEM (discrete element method) \cite{cundall1979discrete} coupled with CFD (computational fluid dynamics) can be used to compute the fluid in an Eulerian framework and each particle in a Lagrangian framework. These methods are known as CFD-DEM and, together with DEM, have been used successfully to simulate in detail the behavior of grains in dense flows \cite{Kloss, Berger, Li, Sun, Sun2, Liu, li2017modeling}.

Reddy and Joshi \cite{Reddy} presented a numerical study on SLFB using an Eulerian two-fluid method. With this method, they investigated the expansion of mono size SLFB for Reynolds numbers going from creeping to turbulent regimes. In addition, Reddy and Joshi \cite{Reddy} investigated binary SLFB for mixing, segregation and layer inversion. The authors compared their results with experiments on relatively large beds (far from the narrow case) and the agreement between them was good. Peng et al. \cite{Peng} studied numerically the segregation of particles in binary SLFB by using a CFD-DEM approach. The simulations were performed in a 0.4 m-high column with a diameter of 0.02 m, and the bed consisted of particles of same size but different densities ($d$ = 1.09 mm and $\rho_p$ = 1600 and 1900 kg/m$^3$). A layer inversion was mimicked in the simulations by placing the high density particles above the low density ones. Peng et al. \cite{Peng} presented the concentration profiles for each species for the steady state regime, i.e., after layer inversion took place, and comparisons with the experimental results of Galvin et al. \cite{Galvin} showed good agreement. The authors also presented the trajectories of one single particle of each species during the layer inversion as well as the steady state. In addition, Peng et al. \cite{Peng} investigated the effect of drag and contact forces on segregation.

Khan et al. \cite{Khan} investigated analytically and numerically how the spatial distribution of phase fractions, segregation and intermixing vary with the dispersion coefficient. For that, they compared different correlations for the dispersion coefficient of binary SLFB and correlated them with the dissipation rate of specific energy. They incorporated some dispersion correlations in the model proposed by Galvin et al. \cite{Galvin} for local volume fractions, and the values obtained agreed well with experimental results. Finally, Khan et al. \cite{Khan} proposed a two-dimensional Eulerian-Eulerian multiphase model that captures the segregation and mixing of binary SLFB.

Abbaszadeh Molaei et al. \cite{Abbaszadeh} investigated numerically the layer inversion in a binary SLFB using the CFD-DEM approach. In their simulations, the bed consisted of glass spheres with $d$ = 0.193 mm and $\rho_p$ = 2510 kg/m$^3$ and carbon spheres with $d$ = 0.778 mm and $\rho_p$ = 1509 kg/m$^3$ fluidized in a 180 mm-high, 10 mm-wide and 1.556 mm-thick box. The authors proposed that layer inversion occurs because, as the liquid velocity increases, the increase in the drag force compensates the decrease in the pressure force for the smaller particles, which does not happen for the larger ones. In addition, they proposed a model that predicts the inversion velocity based on a force balance. The model suited their experimental data better than previous ones.

The use of narrow beds, for which the ratio between the tube and grain diameters is relatively small (of the order of 10), is gaining interest for engineering applications. One example is fluidized-bed bioreactors for biological wastewater treatment, which have a huge potential of employment for both large-scale and domestic treatment of wastewater \cite{Nelson}. Those bioreactors can not only reduce costs but also make wastewater treatment accessible where it does not exist today. The bed may be narrow in those reactors due to the system dimensions and the formation of clusters of bonded particles.

In the case of narrow tubes, confinement effects caused by the influence of walls may affect the process of layer inversion. Zenit et al. \cite{Zenit} measured the collisional particle pressure in SLFB and gravity-driven flows in 50.8 and 106.1 mm-ID tubes, where the bed consisted of  plastic, glass and steel beads of different diameters. The ratio between the tube and particle diameters was within 8 and 52, and, in the specific case of SLFB, Zenit et al. \cite{Zenit} quantified the granular pressure. The authors found that the granular pressure depends on the particle fraction in a non-monotonic way, with a maximum at particle fractions between 30 and 40 $\%$. Zenit et al. \cite{Zenit} proposed that this behavior occurs because the collisions are less frequent for lower particle fractions, whereas they have less impulse for higher fractions. Zenit and Hunt \cite{Zenit2} measured the fluctuation component of grain velocities for almost the same cases as in Zenit et al. \cite{Zenit}. The authors found that the root mean square of the fluctuations varies non-monotonically with the particle fraction, reaching a maximum for particle fractions between 30 and 45 $\%$. Aguilar-Corona et al. \cite{Aguilar} investigated experimentally the collision frequency and the coefficient of restitution of solid particles in SLFB in a narrow tube. For a ratio of 13.33 between the tube and particle diameters, the authors found that the collision frequency is an increasing function of particle fraction and that the normal restitution coefficient varies with the Stokes number.

Recently, C\'u\~nez and Franklin \cite{Cunez} investigated numerically and experimentally the dynamics of granular plugs appearing in SLFB in narrow tubes. The beds consisted of alumina beads with $d$ = 0.6 mm and $\rho_p$ = 3690 kg/m$^3$ fluidized by water flowing in a 25.4 mm-ID tube; therefore, the ratio between the tube and grain diameters was 4.23. For superficial velocities of 0.137 and 0.164 m/s, C\'u\~nez and Franklin \cite{Cunez} showed the existence of granular plugs that propagated upwards, for which they measured the characteristic lengths and celerities. Moreover, based on the network of contact forces obtained by numerical simulations, they showed that confinement is important for the formation of granular plugs.

There are few studies on narrow SLFB, and none of them on layer inversion. Better knowledge of the behavior of individual grains during layer inversion in narrow pipes can enhance our understanding on the effects of walls and confinement on inversion, and, more generally, on the inversion processes itself. Therefore, it can improve engineering applications such as solid classifiers and biological reactors.

This paper investigates experimentally and numerically the dynamics of solid particles during layer inversion in binary SLFB in narrow tubes. Confinement effects caused by the narrow tube, that may give rise to structures such as arches and plugs, change the way in which individual particles move. Although the layer inversion in SLFB has been studied over the last decades, no one investigated it experimentally or numerically in the narrow case. In addition, to the authors' knowledge, detailed experiments on the particles trajectories during the inversion were not reported in previous papers. The objectives of the study are: (i) to investigate how the wall and confinement affect the particle trajectories and concentrations during the layer inversion; (ii) to propose a characteristic time for layer inversion taking into account the effects of confinement. These issues were not addressed in previous papers and are presented here for the first time.

The fluidized beds were formed in a 25.4 mm-ID pipe and consisted of two different solid species: alumina beads with $d$ = 6 mm and $\rho_p$ = 3690 kg/m$^3$, and aluminum beads with $d$ = 4.8 mm and $\rho_p$ = 2760 kg/m$^3$. The ratios between the tube and grain diameters were of 4.23 and 5.29, respectively, corresponding to a very narrow case. The species were chosen in order to force an inversion of layers, mimicking the layer inversion mechanism in the beginning of each run. For that, the initial beds were formed with the lighter particles on the bottom and the heavier ones on the top, for both the experiments and simulations. Afterward, water flows with superficial velocities of 0.137 or 0.164 m/s were imposed in the pipe and the inversion of layers took place. In the experiments, we filmed the fluidized bed with a high-speed camera and automatically identified and tracked individual beads along images by using numerical scripts. In the numerical simulations, we computed the bed evolution with a coupled CFD-DEM code, from which we evaluated the instantaneous positions of each individual particle. The fluid flow is computed with the open source code OpenFOAM, based on FVM (finite volume method), while the granular dynamics is computed with the open source code LIGGGHTS \cite{Kloss, Berger}, which is based on DEM, and both are linked via the open source code \mbox{CFDEM} \cite{Goniva}. Different from previous numerical works on layer inversion, we considered the virtual mass force, which may be important in the case of liquids. We obtained the distances traveled by individual particles during the inversion and the variation of concentrations of each species. In addition, based on our results, we propose a characteristic time for layer inversion. Those findings represent a significant step toward understanding the layer inversion problem.

The next sections present the experimental setup, model description, numerical setup, experimental and numerical results, and the conclusions.

\section{Experimental setup}

The experimental setup consisted of a water tank, a heat exchanger, a centrifugal pump, a flow meter, a flow homogenizer, a 25.4 mm-ID tube vertical section, a 25.4 mm-ID tube horizontal section, and a return line. The granular bed was formed in the vertical section, and the water flowed in a closed loop following the order described in the previous sentence. The vertical section was a 1.2 m-long, 25.4 mm-ID transparent PMMA (Polymethyl methacrylate) tube, of which 0.65 m was the test section, and it was vertically aligned within $\pm 3^{\circ}$. A visual box filled with water was placed around the test section in order to minimize image distortions.

\begin{figure}[ht]
	\centering
	\includegraphics[width=0.9\columnwidth]{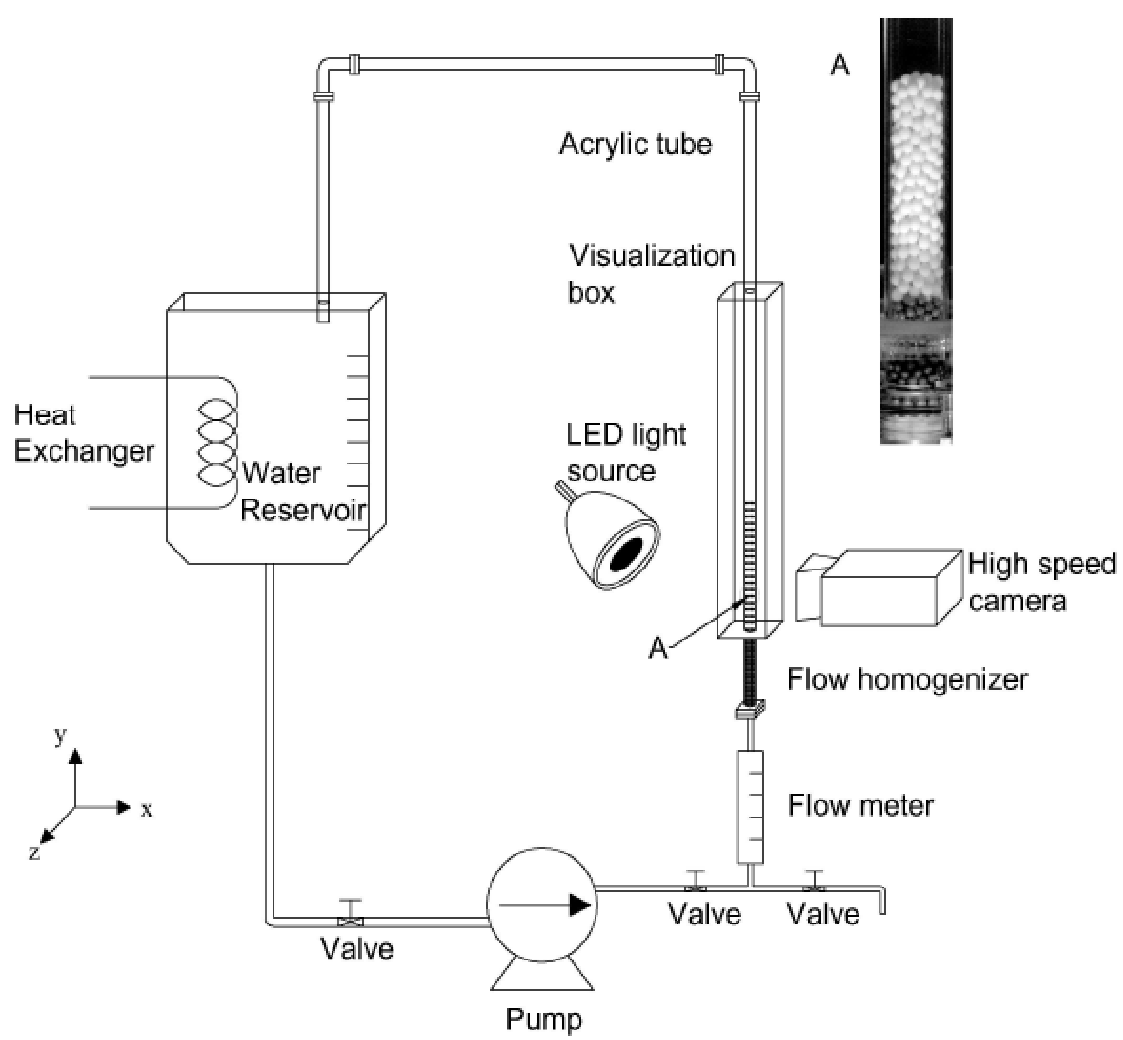}
	\caption{Layout of the experimental setup. The insert shows a photograph of the granular bed in the test section.}
	\label{fig:1}
\end{figure}

\begin{figure}[ht]
	\centering
	\includegraphics[width=0.4\columnwidth]{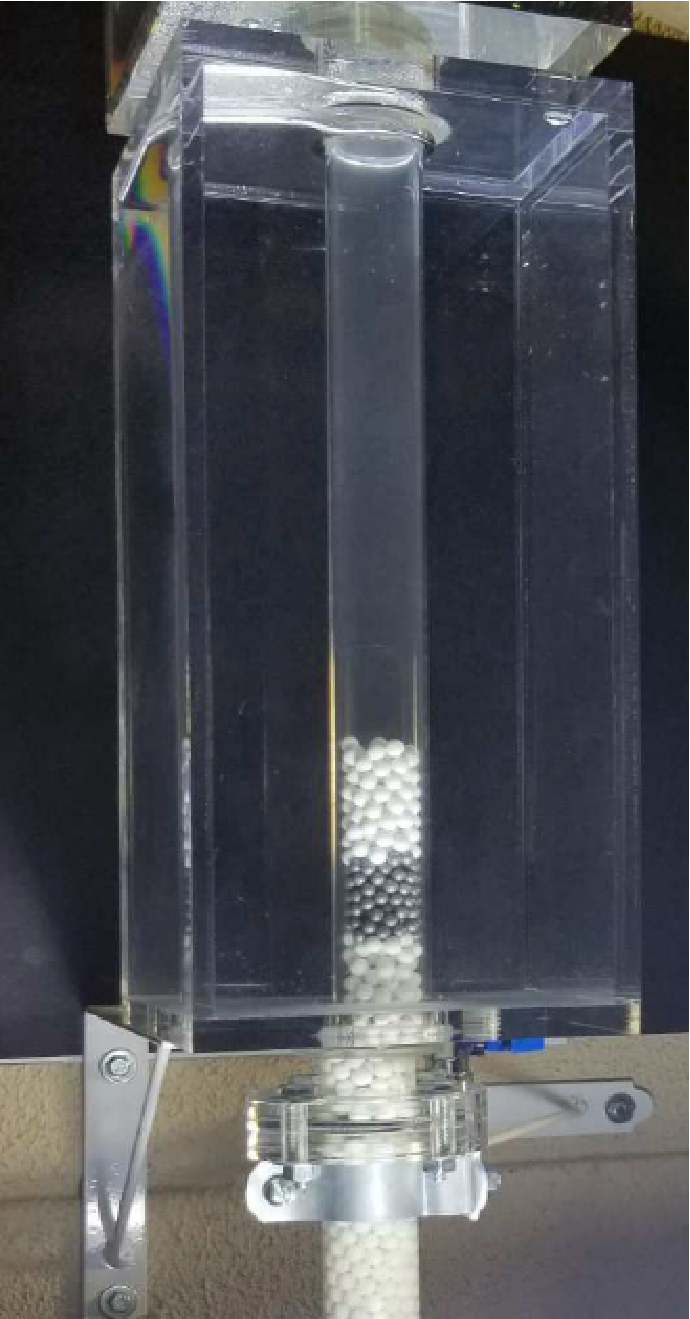}
	\caption{Test section.}
	\label{fig_test_section}
\end{figure}

The centrifugal pump could impose a water flow rate from 0 to 4100 l/h, which was adjusted by a set of valves and by controlling the pump rotation. In order to assure uniform water flows at the test section inlet, a flow homogenizer was placed upstream the test section. It consisted of a 150 mm-long tube containing packed beads with $d$ = 6 mm between fine wire screens. The heat exchanger assured water temperatures within 25$^{\circ}$C $\pm$ 3$^{\circ}$C. Fig.\ref{fig:1} shows the layout of the experimental setup and Fig. \ref{fig_test_section} shows a photograph of the test section.

For the beds, we used alumina beads with $d_1$ = 6 mm $\pm$ 0.03 mm and $S_1$ = $\rho_{p1} / \rho_f$ = 3.69 (species 1), and aluminum beads with $d_2$ = 4.8 mm $\pm$ 0.03 mm and $S_2$ = $\rho_{p2} / \rho_f$ = 2.76 (species 2), where $\rho_f$ the density of the fluid, and the subscripts 1 and 2 correspond to the alumina and aluminum beads, respectively. Four different beds were arranged, each one consisting of 150 beads of each species, 150 beads of species 1 and 250 of species 2, 250 beads of species 1 and 150 of species 2, and 250 beads of each species. The ratios between the tube, $D$, and grain diameters were of 4.23 and 5.29 for species 1 and 2, respectively, and the numbers of Stokes $St_t \,=\, v_t d \rho_p / (9\mu_f)$ and Reynolds $Re_t \,=\, \rho_f  v_t d / \mu_f$ based on terminal velocities, where $v_t$ is the terminal velocity of one single particle and $\mu_f$ is the dynamic viscosity of the fluid, are shown in Tab. \ref{table:table1}. Defined in that way, $St_t$ is the ratio between the characteristic time of solid particles and that of the fluid flow (at rest as a whole) around the solid particle. In its turn, $Re_t$ is the ratio between fluid inertia and viscous effects of the flow around a free falling particle. The relatively high values of $St_t$ and $Re_t$ in Tab. \ref{table:table1} indicate that the particles have considerable inertia with respect to the employed fluid, and that the flow around falling particles induces a wake region. The bed heights for each species at the inception of fluidization were in average 64 and 104 mm ($h_{mf1}$) for the beds consisting of 150 and 250 particles of species 1, respectively, and 30 and 51 mm ($h_{mf2}$) for the beds consisting of 150 and 250 particles of species 2, respectively. From $h_{mf}$, the liquid volume fraction at the inception of fluidization $\varepsilon_{mf}$ was computed for each species. The settling velocity at the inception of fluidization, computed based on the Richardson--Zaki correlation, $v_s = v_t\varepsilon_{mf}^{2.4}$, was 0.11 and 0.06 m/s for species 1 and 2, respectively. Two water flow rates were imposed, $Q$ = 250 l/h and $Q$ = 300 l/h, for which the corresponding superficial velocities $\overline{U} = 4Q/\pi D^2$, fluid velocities through the packed bed $U_f=\overline{U} / \varepsilon_{mf}$, Reynolds numbers based on the tube diameter $Re_D \,=\, \rho_f  \overline{U} D / \mu_f$, and Reynolds numbers based on the grain diameter $Re_d \,=\, \rho_f  \overline{U} d / \mu_f$ are summarized in Tab. \ref{table:table1}. The values of $Re_D$ and $Re_d$ presented in Tab. \ref{table:table1} correspond to a turbulent base flow for the liquid and indicate the presence of wake regions around the solid particles.

\begin{table}[ht]
\caption{Grain diameter $d$, particle type, density ratio $S$, terminal Reynolds number $Re_t$, terminal Stokes number $St_t$, water flow rate $Q$, superficial velocity $\overline{U}$, Reynolds number based on the tube diameter $Re_D$, Reynolds number based on the grain diameter $Re_d$, settling velocity $v_s$, and fluid velocities through the packed bed $U_f$.}
\label{table:table1}
\centering
\begin{tabular}{c c c c c c c c c c c}  
\hline\hline
$d$ & Species & $S$ & $Re_t$  & $St_t$ & $Q$ & $\overline{U}$ & $Re_D$ & $Re_d$ & $v_s$ & $U_f$\\
mm & $\cdots$ & $\cdots$ & $\cdots$ & $\cdots$ & l/h & m/s & $\cdots$ & $\cdots$ & m/s & m/s\\ [0.5ex] 
\hline 
6.0 & 1 & 3.69 & 4026 & 1650 & 250 & 0.137 & 3481 & 822 & 0.11 & 0.29\\
6.0 & 1 & 3.69 & 4026 & 1650 & 300 & 0.164 & 4177 & 987 & 0.11 & 0.35\\
4.8 & 2 & 2.76 & 2269 & 696 & 250 & 0.137 & 3481 & 658 & 0.06 & 0.31\\
4.8 & 2 & 2.76 & 2269 & 696 & 300 & 0.164 & 4177 & 789 & 0.06 & 0.38\\
\hline
\hline 
\end{tabular}
\end{table}

The fluidized bed was filmed with a high-speed camera of CMOS (Complementary Metal Oxide Semiconductor) type with a resolution of 1600 px $\times$ 2560 px at frequencies up to 1400 Hz. LED (low emission diode) lamps were branched to a continuous current source to provide the necessary light while avoiding beating between the light and the camera frequency. In all the experiments, the camera frequency was set to between 100 Hz and 200 Hz and the ROI (region of interest) was fixed to 248 px $\times$ 2560 px. The total fields measured 41 mm $\times$ 420 mm, corresponding to areas of the order of 36 px $\times$ 36 px for each grain of species 1 and 29 px $\times$ 29 px for each grain of species 2.

\section{CFD-DEM equations}
\label{section_model}
The numerical investigation of this study was conducted with {CFD-DEM}. We present next the main equations used in our {CFD-DEM} simulations.

\subsection{Liquid}
The fluid is the continuous phase, being computed in an Eulerian frame. When in the presence of solid particles, one common approach is to use the averaged incompressible Navier-Stokes equations for two-phase flows, whose mass and momentum equations are given by Eqs. \ref{mass} and \ref{qdm}, respectively.

\begin{equation}
{\frac{\partial{\rho_{f}\varepsilon_{f}}}{\partial{t}}+\nabla\cdot(\rho_{f}\varepsilon_{f}\vec{u}_{f})=0}
\label{mass}
\end{equation}

\begin{equation}
\frac{\partial{\rho_{f}\varepsilon_{f}\vec{u}_{f}}}{\partial{t}} + \nabla \cdot (\rho_{f}\varepsilon_{f}\vec{u}_{f}\vec{u}_{f}) = -\varepsilon_{f}\nabla P + \varepsilon_{f}\nabla\cdot \vec{\vec{\tau}}_{f} + \varepsilon_{f}\rho_{f}\vec{g}+\frac{\vec{F}_{pf}}{V_{cell}}
\label{qdm}
\end{equation}

\noindent In Eqs. \ref{mass} and \ref{qdm}, $P$ is the pressure, $\vec{g}$ is the acceleration of gravity, $\vec{\vec{\tau}}_{f}$ is the stress tensor, $\vec{u}_{f}$ and $\varepsilon_{f}$ are, respectively, the mean velocity and volume fraction of the fluid phase, $V_{cell}$ is the volume of the considered cell, and $\vec{F}_{pf}$ is the reaction on the fluid of the drag force on particles and represents the momentum transfer from the fluid to the solids \cite{li2017modeling}, given by:

\begin{equation}
{\vec{F}_{pf}=-\sum_{\forall p \in cell}\frac{V_{p}\beta}{1-\varepsilon_{f}}\left(\vec{u}_{fp} - \vec{u}_{p} \right)}
\label{Ffp}
\end{equation}

\noindent where $\vec{u}_{p}$ is the particle velocity, $V_{p}$ is the particle volume, $\vec{u}_{fp}$ is the liquid velocity at the particle position, and $\beta$ is the drag force coefficient (momentum transfer between phases), given by the Gidaspow model \cite{gidaspow1994multiphase}:

\begin{equation}
\beta=\left\{
\begin{array}{cc}
\frac{3}{4}C_{d}\frac{\rho_{f}\varepsilon_{f}\left(1-\varepsilon_{f}\right)\left|\vec{u}_{f}-\vec{u}_{p}\right|}{d}\varepsilon_{f}^{-2.65} & \left(1-\varepsilon{f}\right) \leq 0.2\\
\\
150\frac{\left(\varepsilon_{f}\right)^{2}\vec{u}_{f}}{\varepsilon_{f}d^{2}}+1.75\frac{\rho_{f}\left(1-\varepsilon_{f}\right)\left|\vec{u}_{f}-\vec{u}_{p}\right|}{d} & \left(1-\varepsilon{f}\right) > 0.2\\
\end{array}\right.
\label{beta}
\end{equation}

\noindent where $C_d$ is a drag coefficient \cite{gidaspow1994multiphase}.

\subsection{Solids}

The solid particles are the disperse phase and are treated as individual objects to be followed in a Lagrangian framework. The dynamics of each particle is computed by the linear and angular momentum equations, given by Eqs. \ref{Fp} and \ref{Tp}, respectively:

\begin{equation}
m_{p}\frac{d\vec{u}_{p}}{dt}= \vec{F}_{fp} + \vec{F}_{press} + \vec{F}_{vm}+ m_p\vec{g} +\sum_{i\neq j}^{N_c} \left(\vec{F}_{c,ij} \right) + \sum_{i}^{N_w} \left( \vec{F}_{c,iw} \right)
\label{Fp}
\end{equation}

\begin{equation}
I_{p}\frac{d\vec{\omega}_{p}}{dt}=\sum_{i\neq j}^{N_c} \vec{T}_{p,ij} + \sum_{i}^{N_w} \vec{T}_{p,iw}
\label{Tp}
\end{equation}

\noindent where, for each individual particle, $m_{p}$ is the mass, $\vec{F}_{fp}$ = $-\vec{F}_{pf}$ is the drag force caused by the fluid, $\vec{F}_{press}$ = $-V_{p}\nabla P + V_{p}\nabla\cdot\vec{\vec{\tau}}_{f}$ = $V_{p}\rho_{f}(D\vec{u}_{f}/Dt -\vec{g})$ is the force due to pressure and stress gradients of the liquid ($D/Dt$ being the material derivative), $\vec{F}_{vm}$ is the virtual mass force, and $\vec{F}_{c,ij}$ and $\vec{F}_{c,iw}$ are the contact forces between particles and between particles and the tube wall, respectively. The term $\vec{T}_{p,ij}$ represents the torque generated by the tangential component of the contact force between particles $i$ and $j$, and $\vec{T}_{p,iw}$ the torque generated by the tangential component of the contact force between particle $i$ and the wall. $I_{p}$ is the moment of inertia, $\vec{\omega}_{p}$ is the angular velocity, $N_c$ - 1 is the number of particles in contact with particle $i$, and $N_w$ is the number of particles in contact with the wall.

In the case of liquids, the viscous and virtual mass forces must be accounted for. In the present simulations, the viscous forces are computed from velocity gradients around the solid particles \cite{Zhou}. The virtual mass force $\vec{F}_{vm}$ is computed by Eq. \ref{Fvm}:

\begin{equation}
\vec{F}_{vm}=0.5\left(1-\varepsilon_{f}\right) V_{p} \rho_{f}\left(\frac{d\vec{u}_{f}}{dt}-\frac{d\vec{u}_{p}}{dt}\right)
\label{Fvm}
\end{equation}

Finally, the contact forces $\vec{F}_{cn,ij}$ are computed using a HSD (Hertzian spring-dashpot) model, for which the normal and tangential components are given by Eqs. \ref{Fcn} and \ref{Fct}, respectively,

\begin{equation}
\vec{F}_{cn,ij}=\left(-k_{n}\delta_{nij}^{3/2}-\eta_{n}\vec{u}_{ij}\cdot\vec{n}_{ij}\right)\vec{n}_{ij}
\label{Fcn}
\end{equation}

\begin{equation}
\vec{F}_{ct,ij}= \left( -k_{t}\delta_{tij} - \eta_{t}\vec{u}_{sij}\cdot\vec{t}_{ij} \right) \vec{t}_{ij}
\label{Fct}
\end{equation}

\noindent where $k$, $\eta$ and $\mu_{fr}$ are the stiffness, damping, and friction coefficients, respectively, which are related to deformation distances in normal, $\delta_n$, and tangential, $\delta_t$, directions. $\vec{n}_{ij}$ is the unit vector connecting the centers of particles with direction from $i$ to $j$, $\vec{t}_{ij} = \vec{u}_{sij}/ \left| \vec{u}_{sij} \right|$, $k_{n}$ and $k_{t}$ are the normal and tangential stiffness coefficients, respectively, $\eta_{n}$ and $\eta_{t}$ are the normal and tangential damping coefficients, respectively, $\vec{u}_{ij}$ is the relative velocity between particle $i$ and particle $j$, and $\vec{u}_{sij}$ is the slip velocity at the contact point. We obtained the values of the stiffness $k$, damping $\eta$ and friction $\mu_{fr}$ coefficients from Tsuji et al. \cite{tsuji1992lagrangian}.

\section{Numerical setup}
\label{section_numerical}

We considered a 0.45 m-long and 25.4 mm-ID vertical tube, filled with alumina spheres with $d_1$ = 6 mm, aluminum spheres with $d_2$ = 4.8 mm, and water. The bed configurations were as in the experiments, consisting of 150 beads of each species, 150 beads of species 1 and 250 of species 2, 250 beads of species 1 and 150 of species 2, and 250 beads of each species. For each bed configuration, water flows corresponding to superficial velocities of $\overline{U}$ = 0.137 and 0.164 m/s were imposed at the tube inlet.

The numerical simulations were performed with the open source code CFDEM \cite{Goniva}, that couples OpenFOAM, computing the fluid motion in an Eulerian frame, to LIGGGHTS, computing the granular dynamics in a Lagrangian frame.  In our simulations, we set OpenFOAM to use the PISO (pressure-implicit with splitting of operators) algorithm, and the particle fraction in each cell was obtained from the particle positions. In addition, because the ratio between the tube and grain diameters is small in the present case, we used the big particle void fraction model of CFDEM (www.cfdem.com), which allows the use of solid particles larger than the CFD cells. In that model, values of void fraction equal to 0 and 1 are associated to solids and fluids, respectively, in the cells whose centers are inside the particle. CFDEM then artificially treats the grains as porous particles by increasing their volume while maintaining the original volume of the solid phase \cite{Kloss2, Mondal}. This allows the use of void fractions in Eqs. \ref{mass} and \ref{qdm}.

A constant time step of $5.0\times 10^{-4}s$ was used for the liquid and a time step of $1.0\times 10^{-5}s$ was used for the solid particles. The density $\rho_f$ and dynamic viscosity $\mu_f$ of the liquid were 1000 kg/m$^3$ and 10$^{-3}$ Pa.s, respectively. The coefficient of restitution $e$ was considered as approximately 0.5 based on the experimental results of Aguilar-Corona et al. \cite{Aguilar}, and the Young's modulus and Poisson ratio were obtained from Ref. \cite{tsuji1992lagrangian}. The friction coefficient was considered as 0.5 based on the experiments of Davin et al. \cite{Davim} and simulations of C\'u\~nez and Franklin \cite{Cunez}. The main parameters used in our simulations are listed in Tab. \ref{tabsim}, where $N$ is the total number of solid particles in the bed.

\begin{table}[ht]
	\begin{center}
	\caption{Parameters used in simulations}
	\begin{tabular}{c c c}
		\hline\hline
		Property & Species 1 & Species 2\\ [0.5ex]
		\hline
		Particle diameter $d$ (mm)  & 6 & 4.8 \\
		Particle density $\rho_p$ (kg/m$^3$) & 3690 & 2760 \\
		Young's Modulus $E$ (GPa) & 300  & 71 \\
		Poisson ratio $\sigma$ & 0.21 & 0.34\\
		Restitution coefficient $e$ & 0.5 & 0.5\\
		Friction coefficient $\mu_{fr}$ & 0.5 & 0.5\\
		Number of particles $N$ for case I & 150 & 150 \\
		Number of particles $N$ for case II & 150 & 250 \\
		Number of particles $N$ for case III & 250 & 150 \\
		Number of particles $N$ for case IV & 250 & 250 \\
		\hline
	\end{tabular}
		\label{tabsim}
    \end{center}
\end{table}

\begin{figure}[ht]
	\centering
	\includegraphics[width=7cm,clip]{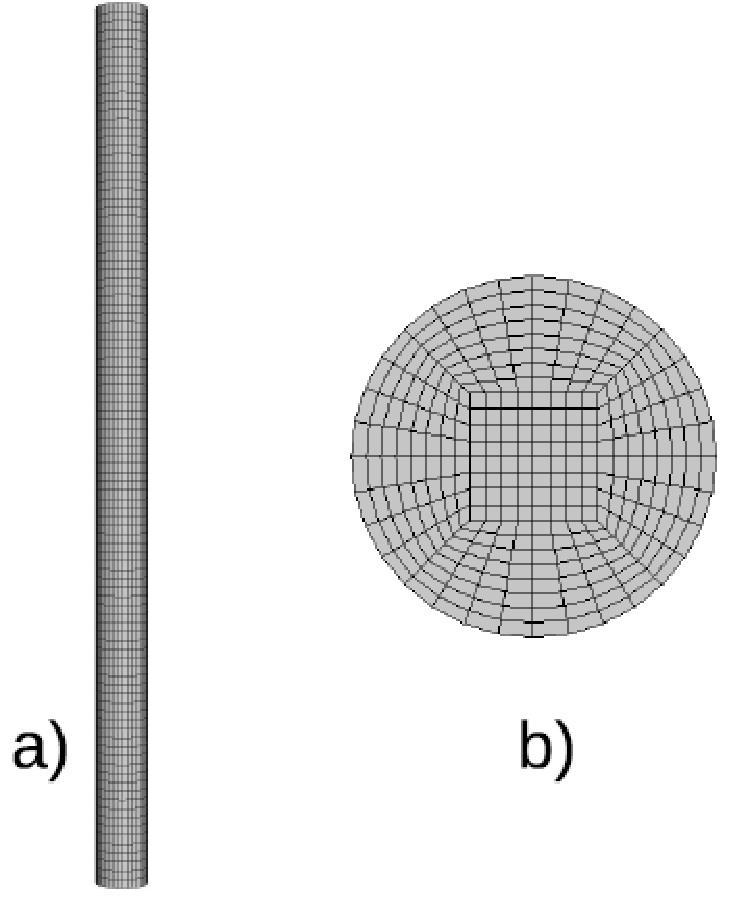}
	\caption{Computational geometry: (a) Side view; (b) Bottom view.}
	\label{mesh}  
\end{figure}

\begin{figure}[ht]
	\centering
	\includegraphics[width=7cm,clip]{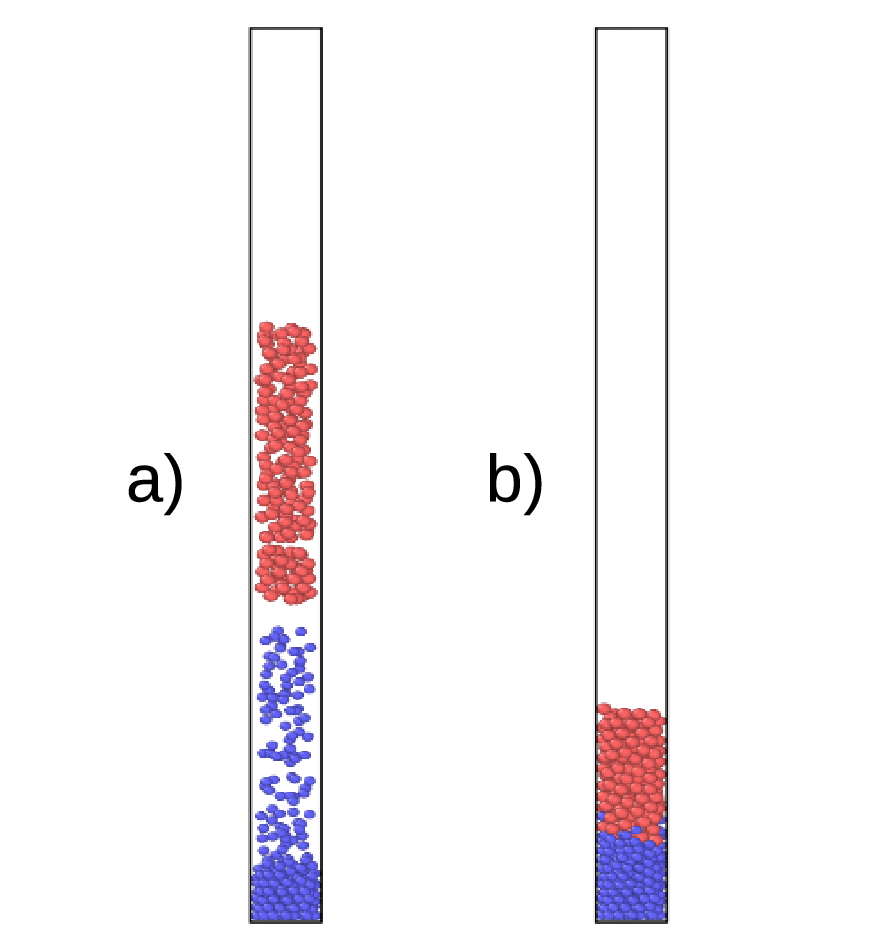}
	\caption{Particle positions: (a) falling particles that were randomly distributed at the initial condition; (b) relaxed state, at the end of initialization. Red color corresponds to species 1 and blue color to species 2.}
	\label{part}  
\end{figure}

For the Eulerian computations, we generated a 0.45 m-long and 25.4 mm-ID vertical cylinder divided by a hexahedral mesh with a total number of 32000 cells, shown in Fig. \ref{mesh}. The tube walls are solid, with no-slip conditions; therefore, the tangential and normal velocities of the liquid were set to zero at the tube walls. At the inlet (bottom boundary), the liquid velocity was set equal to the superficial velocity $\overline{U}$ in the vertical direction and the particles velocities were set to zero. At the outlet (upper boundary), the liquid pressure was specified and the velocity gradient of the liquid was set to zero. As initial conditions, the water flow was set to zero velocity and the particles fell freely in water. After a characteristic time for energy dissipation, the bed reached a stagnant state. The initial conditions are shown in Fig. \ref{part}. Once the stagnant state was reached, the water flow was turned on and the bed was fluidized. 

\section{Results}

\subsection{Experiments}
\label{section_experimental_results}

We observed an initial transient at the beginning of test runs, for all tested conditions. The transient consisted of the upward displacement of the entire bed as a single granular plug, with beads at its bottom falling freely in the ascending liquid. In this way, as the initial plug moved upward its length decreased, the initial transient finishing when all the particles have left the initial plug by free fall. Once particles of both species have reached the bottom of the tube, with the aluminum beads still below, they started mixing with each other even before the initial plug have disappeared, initiating the inversion of layers. At the same time, a few aluminum beads within the ascending plug migrate toward its top. The initial transient characterized by the ascending plug is the result of a blockage due to arches within the granular plug, as shown in Subsection \ref{section_numerical_results} in terms of the network of contact forces. To the authors' knowledge, this is the first time that it is presented, previous works on layer inversion not showing it. Therefore, the transient plug is caused by high confinement, being characteristic of the very narrow case. The initial transient and the bed dynamics following it are shown in a movie from one experiment and an animation from one numerical simulation attached to this paper as Supplementary Material \cite{Supplemental}.

\begin{figure}[ht]
		\begin{minipage}{0.5\linewidth}
			\begin{tabular}{c}
				\includegraphics[width=0.90\linewidth]{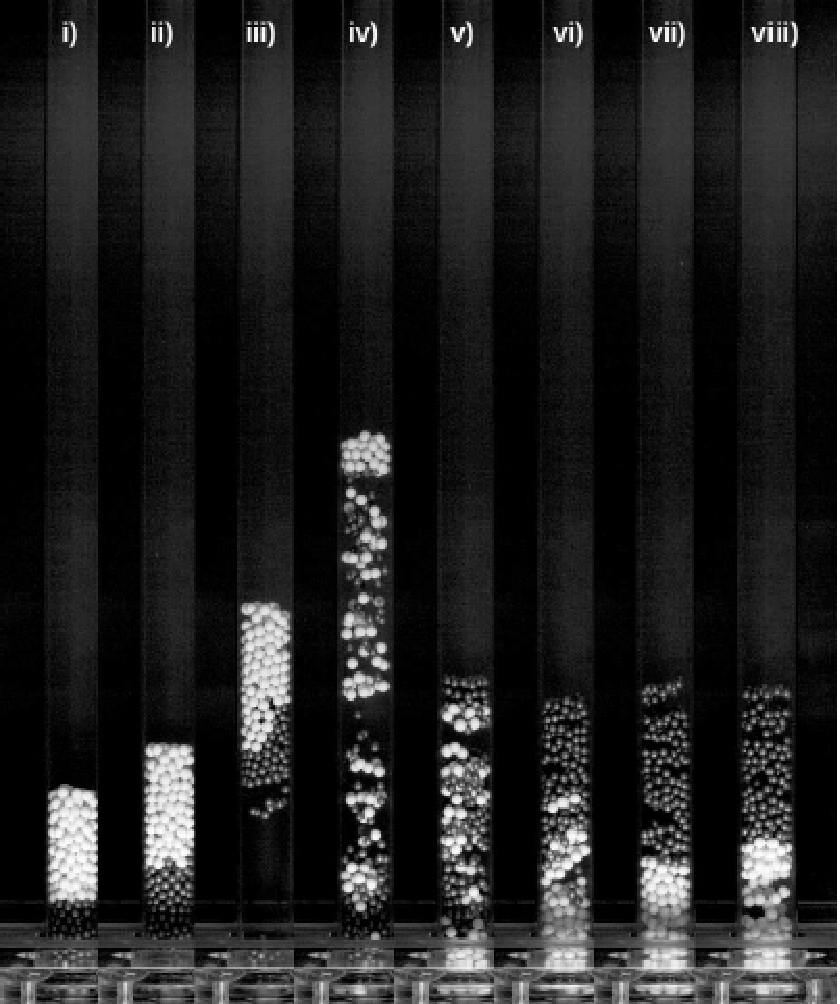}\\
				(a)
			\end{tabular}
		\end{minipage}
		\hfill
		\begin{minipage}{0.5\linewidth}
			\begin{tabular}{c}
				\includegraphics[width=0.90\linewidth]{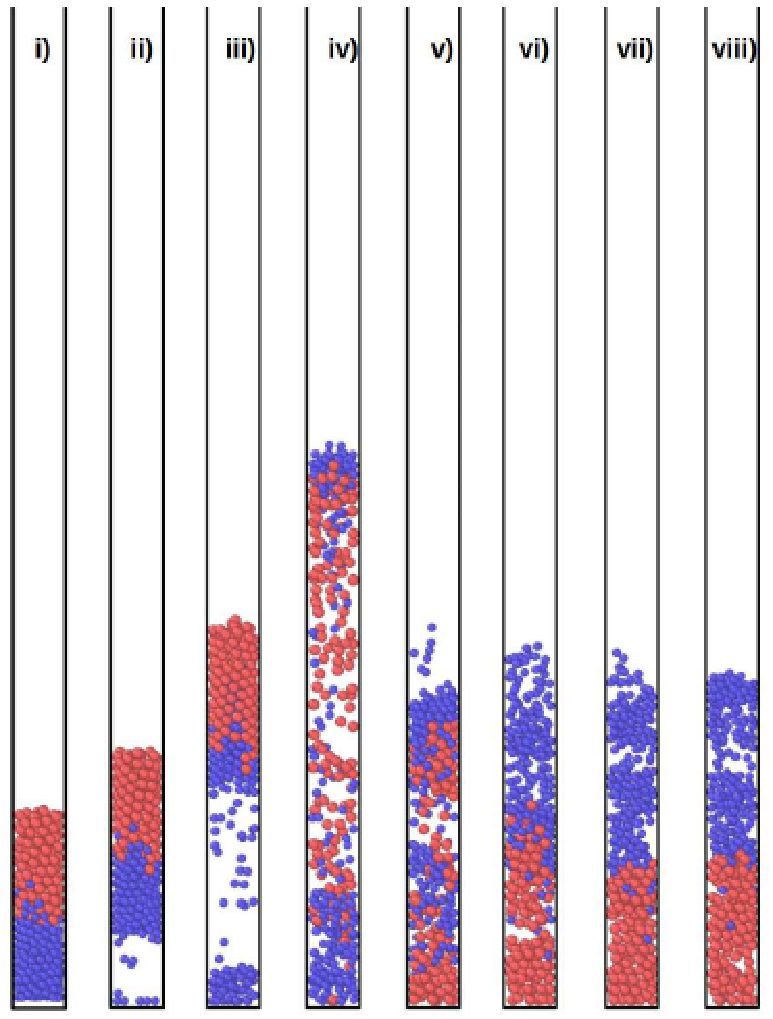}\\
				(b)
			\end{tabular}
		\end{minipage}
		\hfill
		\caption{Instantaneous snapshots of particle positions for case II and $\overline{U}$ = 0.137 m/s. (a) Experiments; (b) numerical simulations. The corresponding times are: (i) 0 s; (ii) 3 s; (iii) 6 s; (iv) 9 s; (v) 12 s; (vi) 15 s; (vii) 18 s; (viii) 21 s.}
		\label{fig:snapshots_150_250_250}
	 \end{figure}

\begin{figure}[ht]
 	\begin{minipage}{0.5\linewidth}
 		\begin{tabular}{c}
 			\includegraphics[width=0.90\linewidth]{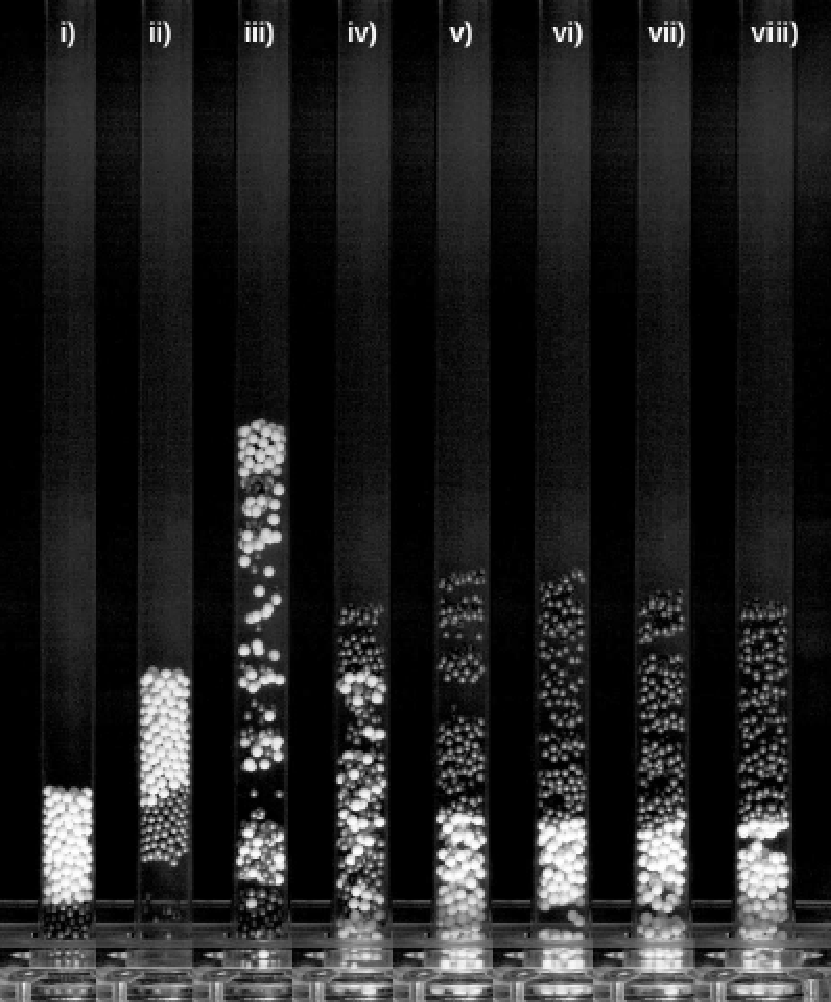}\\
 			(a)
 		\end{tabular}
 	\end{minipage}
 	\hfill
 	\begin{minipage}{0.5\linewidth}
 		\begin{tabular}{c}
 			\includegraphics[width=0.90\linewidth]{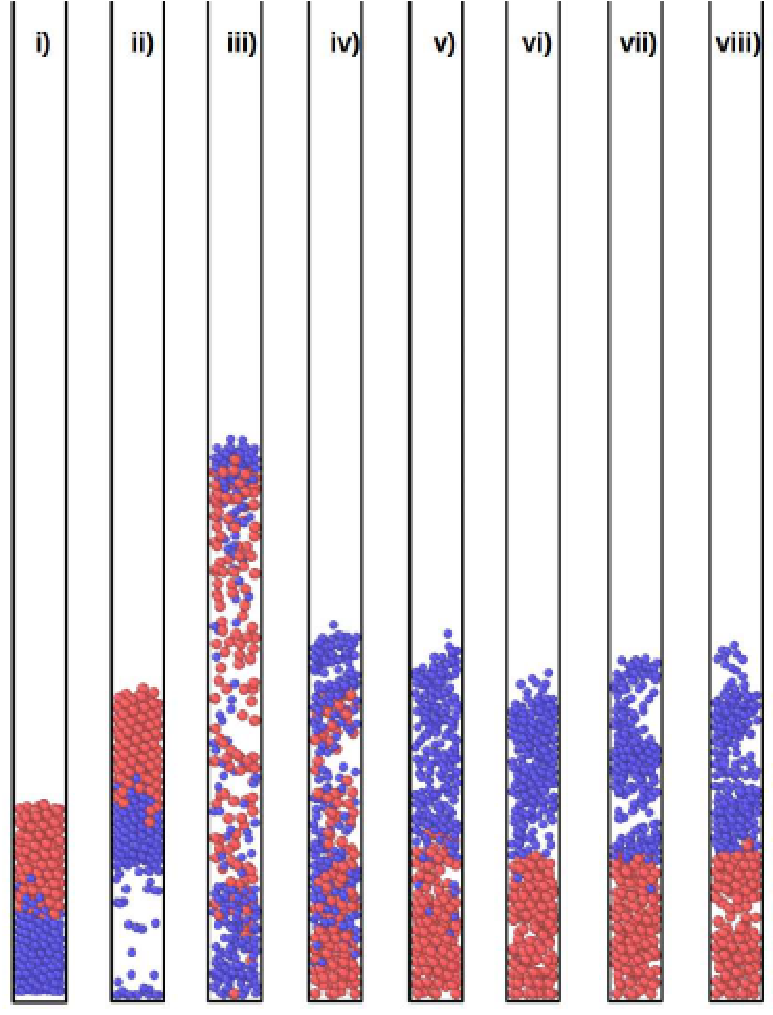}\\
 			(b)
 		\end{tabular}
 	\end{minipage}
	\hfill
	\caption{Instantaneous snapshots of particle positions for case II and $\overline{U}$ = 0.164 m/s. (a) Experiments; (b) numerical simulations. The corresponding times are: (i) 0 s; (ii) 3 s; (iii) 6 s; (iv) 9 s; (v) 12 s; (vi) 15 s; (vii) 18 s; (viii) 21 s.}
	\label{fig:snapshots_150_250_300}
 \end{figure}

When all the particles have left the initial plug and fell to the bottom, the inversion process continues. From this time on, the mixing increases and then decreases, and eventually the layers are inversed. During the inversion of layers, small granular clusters are formed within the bed and when the inversion is almost completed these clusters give origin to granular plugs. The formation of those structures is the result of a modulation caused by the fluid drag, virtual mass and viscous forces, gravity, friction between grains and between grains and the walls, and collisions between grains and between grains and the walls. The granular structures during the inversion can be seen in Figs. \ref{fig:snapshots_150_250_250} and \ref{fig:snapshots_150_250_300}, which show instantaneous snapshots of particle positions for case $II$ for both experiments and numerical simulations (the latter presented in Subsection \ref{section_numerical_results}). The corresponding fluid velocity and times are in the caption of figures. Cases $I$, $III$, and $IV$ present similar behaviors (the respective figures can be seen in the  Supplementary Material \cite{Supplemental}).

\begin{figure}[ht]
	\centering
	\includegraphics[width=0.6\columnwidth]{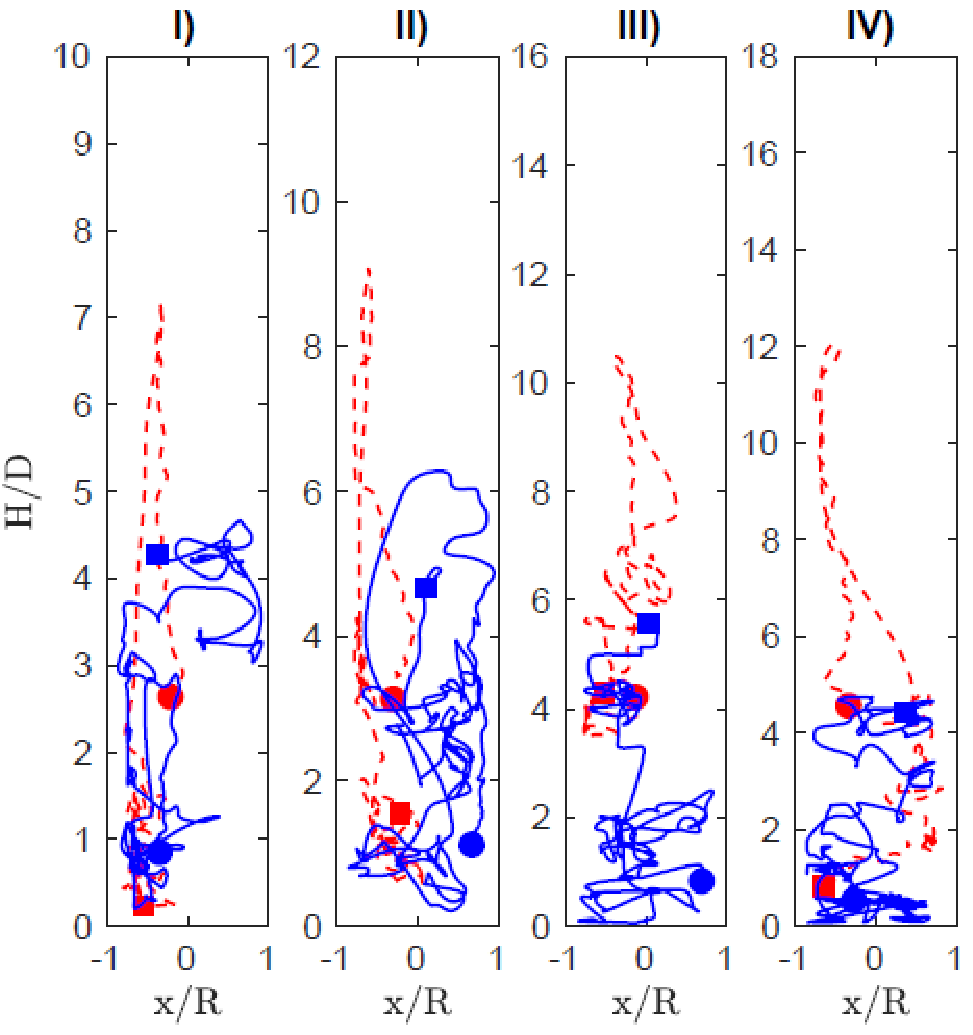}
	\caption{Equivalent trajectories obtained from experiments of one particle of each species from the beginning to the end of inversion. The dashed red line corresponds to species 1 and the continuous blue line to species 2. Circles and squares correspond to initial and final positions, respectively. From left to right, graphics are related to cases $I$, $II$, $III$ and $IV$, in this order, and $\overline{U}$ = 0.137 m/s.}
	\label{fig_tracking_exp250}
\end{figure}

\begin{figure}[ht]
	\centering
	\includegraphics[width=0.6\columnwidth]{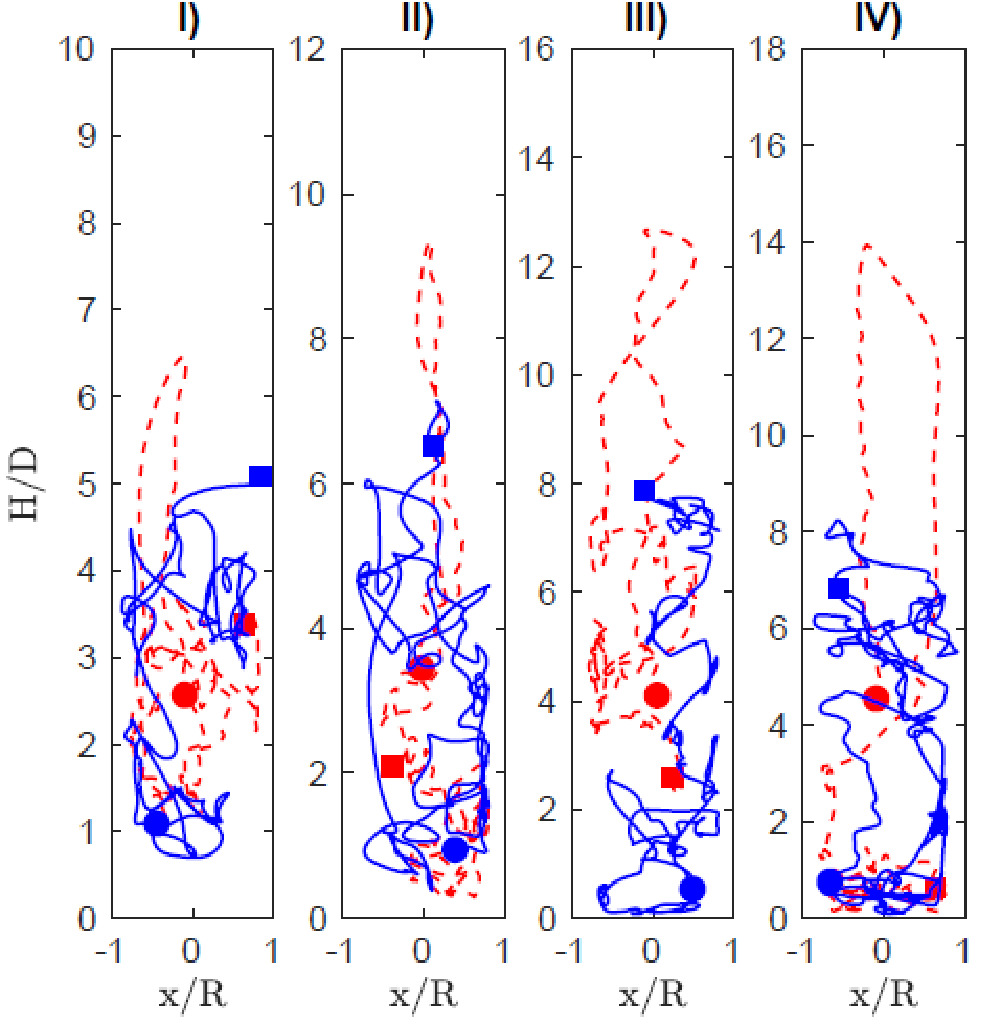}
	\caption{Equivalent trajectories obtained from experiments of one particle of each species from the beginning to the end of inversion. The dashed red line corresponds to species 1 and the continuous blue line to species 2. Circles and squares correspond to initial and final positions, respectively. From left to right, graphics are related to cases $I$, $II$, $III$ and $IV$, in this order, and $\overline{U}$ = 0.164 m/s.}
	\label{fig_tracking_exp300}
\end{figure}

During the inversion process, particles of each species follow typical paths, oscillating between the bottom and top regions of the bed, with the traveled vertical distances decreasing with time. By the end of the process, the aluminum particles oscillate within the top region and the alumina particles within the bottom region of the bed. We identified and followed some beads of both types in the high-speed movies by using numerical scripts developed in the course of this study. These scripts are available at Ref. \cite{Supplemental3}. Figs. \ref{fig_tracking_exp250} and \ref{fig_tracking_exp300} show the equivalent trajectories of one particle of each type for $\overline{U}$ = 0.137 m/s and $\overline{U}$ = 0.164 m/s, respectively, for cases $I$, $II$, $III$ and $IV$, from the beginning to the end of inversion. The dashed red line corresponds to species 1, the continuous blue line to species 2, and circles and squares indicate, respectively, the initial and final positions. In the graphics, the vertical axis corresponds to the height normalized by the tube diameter, $H/D$, and the horizontal axis to the lateral distance normalized by the tube radius, $x/R$. Given the three-dimensional nature of the motion, trajectories of individual particles were difficult to obtain for long times because the camera had a frontal point of view. For this reason, Figs. \ref{fig_tracking_exp250} and \ref{fig_tracking_exp300} show some displacements of different particles in the frontal part of the bed along time in order to reconstruct a typical motion by parts, which we call here the equivalent trajectory. In the case of numerical simulations, presented in Subsection \ref{section_numerical_results}, this was not necessary because the numerical results contain the instantaneous positions of each individual particle.

After inversion took place, the vertical coordinates of the top of the bed and the interface between species 1 and 2 oscillated around mean values, as can be seen in Supplementary Material \cite{Supplemental}. From the images, we computed the mean values of the vertical coordinates of both the top of the bed and the interface between species 1 and 2, normalized by the tube diameter, $z_{top}/D$ and $z_{int}/D$, respectively, after inversion took place. At the initial conditions, these values are given, respectively, by $z_{top,ini}/D$ and $z_{int,ini}/D$. In addition, we computed the standard deviations of $z_{top}$ and $z_{int}$, $\sigma_{ztop}/D$ and $\sigma_{zint}/D$, respectively. Tab. \ref{table:table_hfinal_exp} presents $z_{top,ini}/D$, $z_{int,ini}/D$, $z_{top}/D$, $z_{int}/D$, $z_{top}$ and $z_{int}$ for cases I to IV. These values are compared with those obtained from numerical simulations in Subsection \ref{section_numerical_results}.

\begin{table}[ht]
\caption{Final heights obtained experimentally. From left to right: case, superficial velocity $\overline{U}$, mean vertical coordinate of the top of the bed normalized by the tube diameter at the initial condition $z_{top,ini}/D$, mean vertical coordinate of the interface between species 1 and 2 normalized by the tube diameter at the initial condition $z_{int,ini}/D$, mean vertical coordinate of the top of the bed normalized by the tube diameter after inversion took place $z_{top}/D$, mean vertical coordinate of the interface between species 1 and 2 normalized by the tube diameter after inversion took place $z_{int}/D$, standard deviation of the mean vertical coordinate of the top of the bed normalized by the tube diameter $\sigma_{ztop}/D$, and standard deviation of the mean vertical coordinate of the interface between species 1 and 2 normalized by the tube diameter $\sigma_{zint}/D$.}
\label{table:table_hfinal_exp}
\centering
\begin{tabular}{c c c c c c c c}  
\hline\hline
Case & $\overline{U}$ & $z_{top,ini}/D$ & $z_{int,ini}/D$ & $z_{top}/D$ & $z_{int}/D$  & $\sigma_{ztop}/D$ & $\sigma_{zint}/D$\\
$\cdots$ & m/s & $\cdots$ & $\cdots$ & $\cdots$ & $\cdots$ & $\cdots$ & $\cdots$\\ [0.5ex] 
\hline 
$I$ & 0.137 & 3.7 & 1.2 & 4.4 & 2.2 & 0.2 & 0.1\\
$I$ & 0.164 & 3.7 & 1.2 & 5.8 & 3.2 & 0.4 & 0.3\\
$II$ & 0.137 & 4.5 & 2.0 & 5.8 & 2.5 & 0.2 & 0.1\\
$II$ & 0.164 & 4.5 & 2.0 & 7.7 & 3.3 & 0.4 & 0.3\\
$III$ & 0.137 & 5.3 & 1.2 & 6.6 & 4.5 & 0.2 & 0.2\\
$III$ & 0.164 & 5.3 & 1.2 & 8.6 & 5.9 & 0.4 & 0.3\\
$IV$ & 0.137 & 6.1 & 2.0 & 7.9 & 4.6 & 0.3 & 0.4\\
$IV$ & 0.164 & 6.1 & 2.0 & 10.3 & 5.9 & 0.4 & 0.4\\
\hline
\hline 
\end{tabular}
\end{table}

The inversion process can also be evaluated from profiles of mean volume fractions of each species at different times. One way to determine it is by measuring the quantity of grains of each species in the pipe cross section as a function of the bed height. However, because we cannot access all the particles within different cross sections of the tube, the volume fractions would need to be estimated from the frontal view of the bed. With this procedure, the accuracy would be compromised. For this reason, and given the high fidelity of our numerical simulations, we prefer to present volume fractions based only on the numerical simulations in Subsection \ref{section_numerical_results}.

\subsection{Numerical Simulations}
\label{section_numerical_results}

We performed numerical simulations using CFD-DEM for the same cases measured experimentally. In this way, the numerical results can be compared directly with experiments. Figs. \ref{fig:snapshots_150_250_250} and \ref{fig:snapshots_150_250_300} show instantaneous snapshots of particle positions for case $II$, for both experiments and numerical simulations. The corresponding times are in the caption of figures. Comparing the particle positions from the numerical simulations with experiments, we note that the initial transient, mixing levels, granular structures and inversion dynamics are predicted correctly by the numerical simulations. Cases $I$, $III$, and $IV$ present similar behaviors (the respective figures can be seen in the  Supplementary Material \cite{Supplemental}).

As for the experiments, we computed $z_{top}/D$, $z_{int}/D$, $z_{top,ini}/D$, $z_{int,ini}/D$, $\sigma_{ztop}/D$ and $\sigma_{zint}/D$, that are presented in Tab. \ref{table:table_hfinal_num} for cases I to IV. The values obtained numerically are in accordance with the experiments. This is shown also in Tab. \ref{table:table_hfinal_num2}, which presents the relative differences between numerical and experimental values for $z_{top}$ and $z_{int}$, $\delta_{ztop}$ and $\delta_{zint}$, respectively. These values were obtained as the absolute difference between the numerical and experimental results for $z_{top}$ and $z_{int}$ divided by the respective experimental value. The larger difference is 16.7 \%, most part of differences being below 6 \%. Therefore, taking into account the uncertainties related to measurements in discrete media, and the fact that the bed oscillated at the end of the inversion of layers, we consider that those differences are reasonable and that the simulations captured well the final bed heights.

\begin{table}[ht]
\caption{Final heights obtained numerically. From left to right: case, superficial velocity $\overline{U}$, mean vertical coordinate of the top of the bed normalized by the tube diameter at the initial condition $z_{top,ini}/D$, mean vertical coordinate of the interface between species 1 and 2 normalized by the tube diameter at the initial condition $z_{int,ini}/D$, mean vertical coordinate of the top of the bed normalized by the tube diameter after inversion took place $z_{top}/D$, mean vertical coordinate of the interface between species 1 and 2 normalized by the tube diameter after inversion took place $z_{int}/D$, standard deviation of the mean vertical coordinate of the top of the bed normalized by the tube diameter $\sigma_{ztop}/D$, and standard deviation of the mean vertical coordinate of the interface between species 1 and 2 normalized by the tube diameter $\sigma_{zint}/D$.}
\label{table:table_hfinal_num}
\centering
\begin{tabular}{c c c c c c c c}  
\hline\hline
Case & $\overline{U}$ & $z_{top,ini}/D$ & $z_{int,ini}/D$ & $z_{top}/D$ & $z_{int}/D$  & $\sigma_{ztop}/D$ & $\sigma_{zint}/D$\\
$\cdots$ & m/s & $\cdots$ & $\cdots$ & $\cdots$ & $\cdots$ & $\cdots$ & $\cdots$\\ [0.5ex] 
\hline 
$I$ & 0.137 & 3.6 & 1.2 & 4.6 & 2.6 & 0.2 & 0.1\\
$I$ & 0.164 & 3.6 & 1.2 & 5.3 & 3.1 & 0.4 & 0.2\\
$II$ & 0.137 & 4.3 & 1.9 & 6.0 & 2.6 & 0.3 & 0.1\\
$II$ & 0.164 & 4.3 & 1.9 & 7.6 & 3.6 & 0.4 & 0.2\\
$III$ & 0.137 & 5.2 & 1.2 & 6.4 & 4.5 & 0.2 & 0.2\\
$III$ & 0.164 & 5.2 & 1.2 & 8.6 & 6.2 & 0.7 & 0.5\\
$IV$ & 0.137 & 5.9 & 1.9 & 7.8 & 4.1 & 0.2 & 0.2\\
$IV$ & 0.164 & 5.9 & 1.9 & 10.1 & 5.7 & 0.4 & 0.7\\
\hline
\hline 
\end{tabular}
\end{table}

\begin{table}[ht]
\caption{Comparison between vertical coordinates after inversion took place obtained experimentally and numerically. From left to right: case, superficial velocity $\overline{U}$, and relative difference between numerical and experimental values for $z_{top}$ and $z_{int}$, $\delta_{ztop}$ and $\delta_{zint}$, respectively.}
\label{table:table_hfinal_num2}
\centering
\begin{tabular}{c c c c}  
\hline\hline
Case & $\overline{U}$ & $\delta_{ztop}$ & $\delta_{zint}$\\
$\cdots$ & m/s & \% & \%\\ [0.5ex] 
\hline 
$I$ & 0.137 & 5.8 & 16.7\\
$I$ & 0.164 & 9.3 & 2.6\\
$II$ & 0.137 & 3.3 & 3.0\\
$II$ & 0.164 & 2.1 & 6.7\\
$III$ & 0.137 & 3.3 & 1.2\\
$III$ & 0.164 & 0.3 & 3.7\\
$IV$ & 0.137 & 1.3 & 9.2\\
$IV$ & 0.164 & 2.3 & 2.9\\
\hline
\hline 
\end{tabular}
\end{table}

\begin{figure}[ht]
	\centering
	\includegraphics[width=0.6\columnwidth]{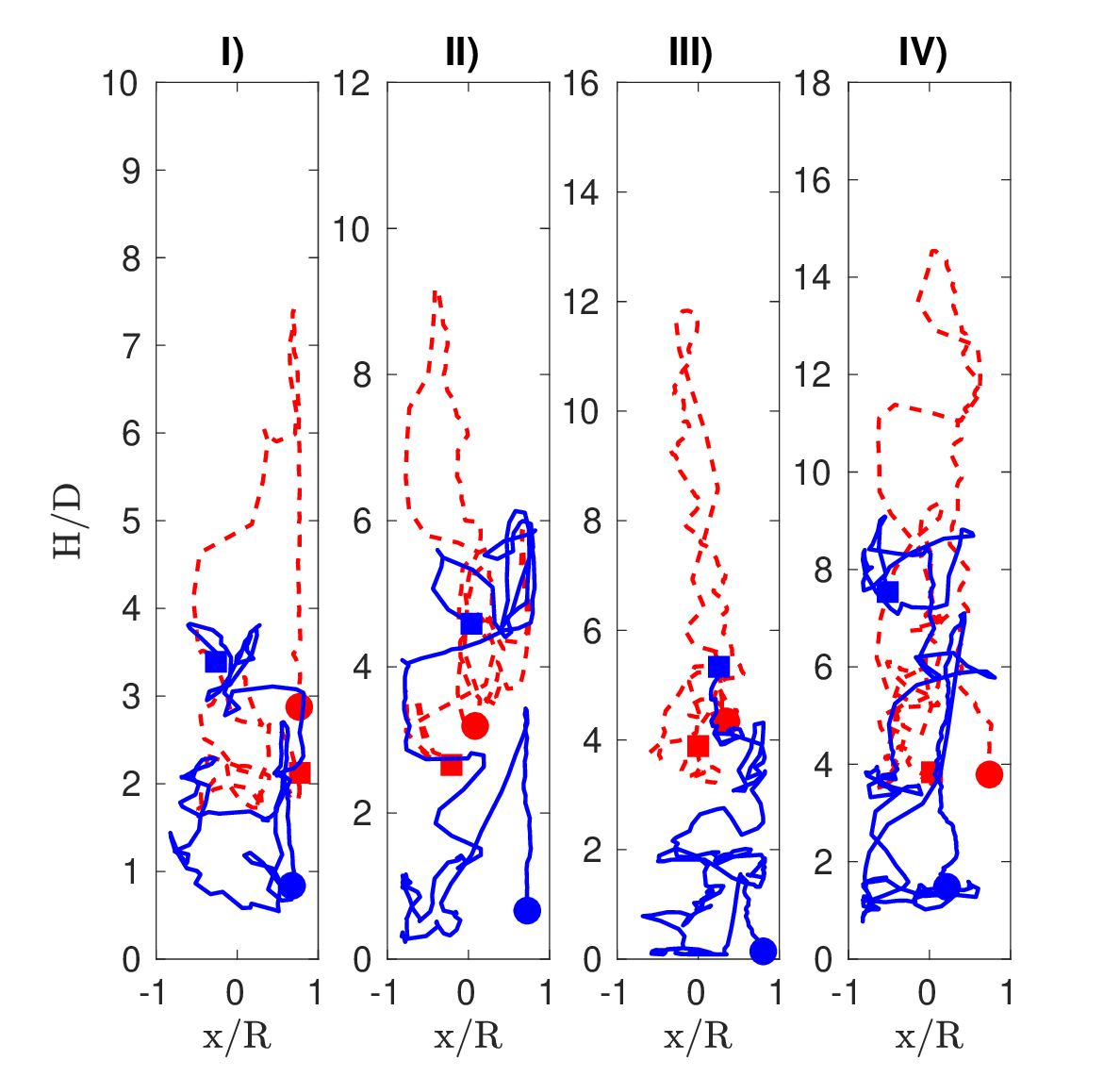}
	\caption{Trajectories obtained numerically for one particle of each species from the beginning to the end of inversion. The dashed red line corresponds to species 1 and the continuous blue line to species 2. Circles and squares correspond to initial and final positions, respectively. From left to right, graphics are related to cases $I$, $II$, $III$ and $IV$, in this order, and $\overline{U}$ = 0.137 m/s.}
	\label{fig_tracking_num250}
\end{figure}

\begin{figure}[ht]
	\centering
	\includegraphics[width=0.6\columnwidth]{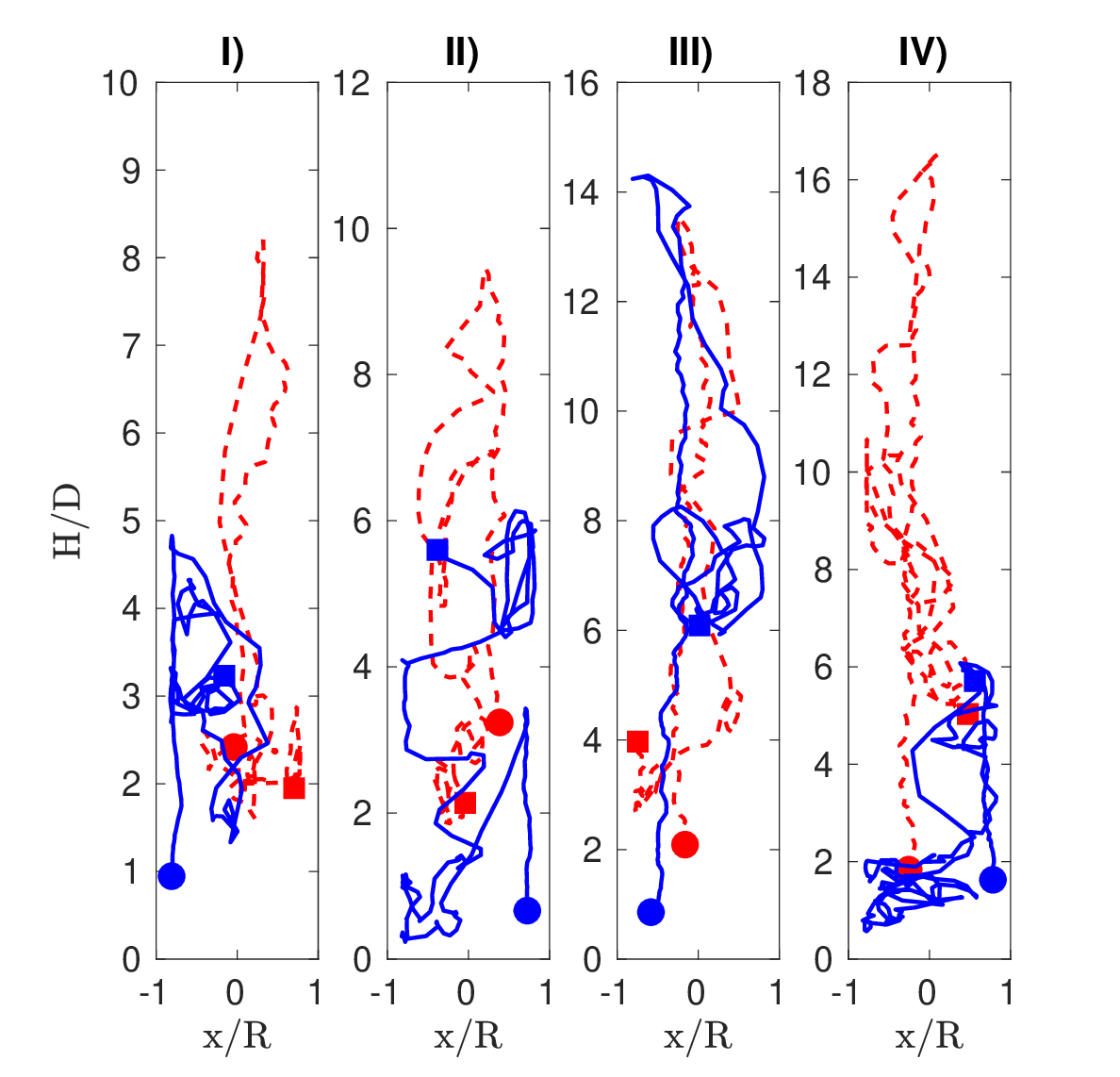}
	\caption{Trajectories obtained numerically for one particle of each species from the beginning to the end of inversion. The dashed red line corresponds to species 1 and the continuous blue line to species 2. Circles and squares correspond to initial and final positions, respectively. From left to right, graphics are related to cases $I$, $II$, $III$ and $IV$, in this order, and $\overline{U}$ = 0.164 m/s.}
	\label{fig_tracking_num300}
\end{figure}

One of the advantages of CFD-DEM simulations with regard to experiments involving grains is the knowledge of the instantaneous positions of each particle. Therefore, the trajectories of individual grains are readily accessible from the numerical results. Figs. \ref{fig_tracking_num250} and \ref{fig_tracking_num300} show the trajectories of one particle of each type for $\overline{U}$ = 0.137 m/s and $\overline{U}$ = 0.164 m/s, respectively, for cases $I$, $II$, $III$ and $IV$ from the beginning to the end of inversion. The dashed red line corresponds to species 1, the continuous blue line to species 2, and circles and squares indicate, respectively, the initial and final positions. As for Figs. \ref{fig_tracking_exp250} and \ref{fig_tracking_exp300}, the vertical axis corresponds to $H/D$ and the horizontal axis to $x/R$. However, different from the equivalent trajectories obtained from experiments, we followed single particles in three dimensions, even though Figs. \ref{fig_tracking_num250} and \ref{fig_tracking_num300} present only a planar view. As for the experimental results (Figs. \ref{fig_tracking_exp250} and \ref{fig_tracking_exp300}), particles of each species follow oscillating paths between the bottom and top regions of the bed, with decreasing traveled distances along time, and with the aluminum and alumina particles oscillating within the top and bottom regions, respectively, by the end of the inversion process. Although Figs. \ref{fig_tracking_num250} and \ref{fig_tracking_num300} present the trajectories of only one particle of each type (for better visualization of trajectories), they represent typical trajectories found in our numerical simulations during the inversion of layers \cite{Supplemental3}.

Based on the instantaneous positions of particles obtained numerically, we computed the total distance traveled by each individual particle from the beginning to the end of inversion. We computed those distances only from numerical data because the experimental ones were constructed by assembling different two-dimensional trajectories; therefore, the trajectories experimentally obtained, although similar to the numerical ones, do not correspond to the three-dimensional motion of a single particle. Mean values and standard deviations were afterward computed by considering all the particles within the bed. Tab. \ref{table:table_num} presents, for species 1 and 2, the mean distances $\Delta$ and standard deviations $\sigma$ normalized by bed heights at the inception of fluidization $h_{mf}$. Subscripts 1 and 2 refer to species 1 and 2, respectively. The choice of $h_{mf}$ for normalization was based on the fact that distances traveled during inversion are proportional to the bed height (dimensional values can be obtained by using $h_{mf}$ values listed in Tab. \ref{table:table_num}). In addition, Tab. \ref{table:table_num} presents $A = t_{inv}/t_{scale}$, which is the ratio between the time for inversion $t_{inv}$ and the timescale $t_{scale} = h_{mf}/\overline{U}$. We note that traveled distances are, roughly, within 5 to 6 times $h_{mf}$ for species 1 and within 6.5 to 7.5 times $h_{mf}$ for species 2, corresponding to traveled distances 25 \% higher for species 2. Larger distances were expected for species 2 because they are smaller, which means that they can move through smaller void regions.

\begin{table}[ht]
\caption{Case, superficial velocity $\overline{U}$, bed height at the inception of fluidization $h_{mf}$, mean values of normalized distances traveled by species 1 and 2, $\Delta_1/h_{mf}$ and $\Delta_2/h_{mf}$, respectively, standard deviation of normalized distances traveled by species 1 and 2, $\sigma_1/h_{mf}$ and $\sigma_2/h_{mf}$, respectively, ratio between traveled distances of species 2 and 1, $\Delta_2/\Delta_1$, and $A = t_{inv}/t_{scale}$.}
\label{table:table_num}
\centering
\begin{tabular}{c c c c c c c c c}  
\hline\hline
Case & $\overline{U}$ & $h_{mf}$ & $\Delta_1/h_{mf}$  & $\sigma_1/h_{mf}$ & $\Delta_2/h_{mf}$ & $\sigma_2/h_{mf}$ & $\Delta_2/\Delta_1$ & $A$\\
$\cdots$ & m/s & mm & $\cdots$ & $\cdots$ & $\cdots$ & $\cdots$ & $\cdots$ & $\cdots$\\ [0.5ex] 
\hline 
$I$ & 0.137 & 90 & 5.4 & 1.0 & 6.8 & 1.0 & 1.27 & 23\\
$I$ & 0.164 & 90 & 5.5 & 1.1 & 7.0 & 1.0 & 1.28 & 16 \\
$II$ & 0.137 & 112 & 5.5 & 1.1 & 6.5 & 1.1 & 1.18 & 22\\
$II$ & 0.164 & 112 & 5.6 & 1.2 & 6.7 & 1.1 & 1.20 & 22\\
$III$ & 0.137 & 129 & 5.0 & 1.1 & 6.5 & 0.7 & 1.29 & 22\\
$III$ & 0.164 & 129 & 5.1 & 1.1 & 6.6 & 0.7 & 1.29 & 19\\
$IV$ & 0.137 & 151 & 6.2 & 1.4 & 7.3 & 0.9 & 1.18 & 22\\
$IV$ & 0.164 & 151 & 6.2 & 1.4 & 7.4 & 1.0 & 1.19 & 20\\
\hline
\hline 
\end{tabular}
\end{table}

\begin{figure}[ht]
 	\begin{minipage}[c]{\columnwidth}
 		\begin{center}
 			\includegraphics[width=0.90\linewidth]{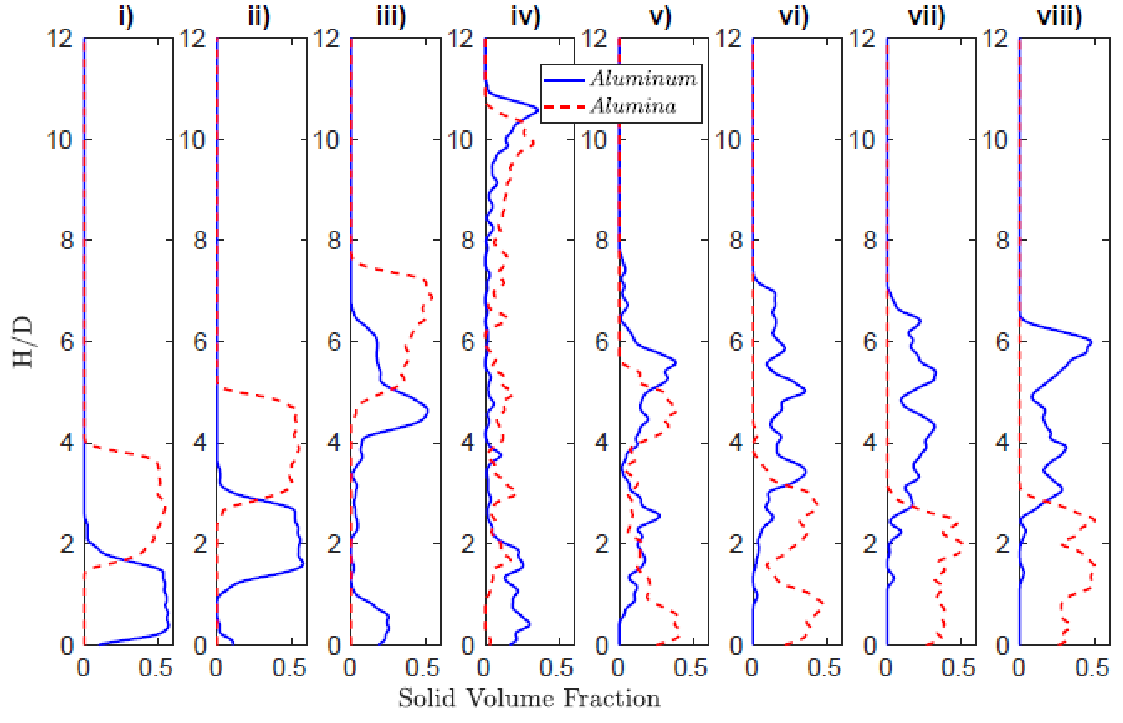}\\
 			(a)
 		\end{center}
 	\end{minipage}
 	\hfill
 	\begin{minipage}[c]{\columnwidth}
 		\begin{center}
 			\includegraphics[width=0.90\linewidth]{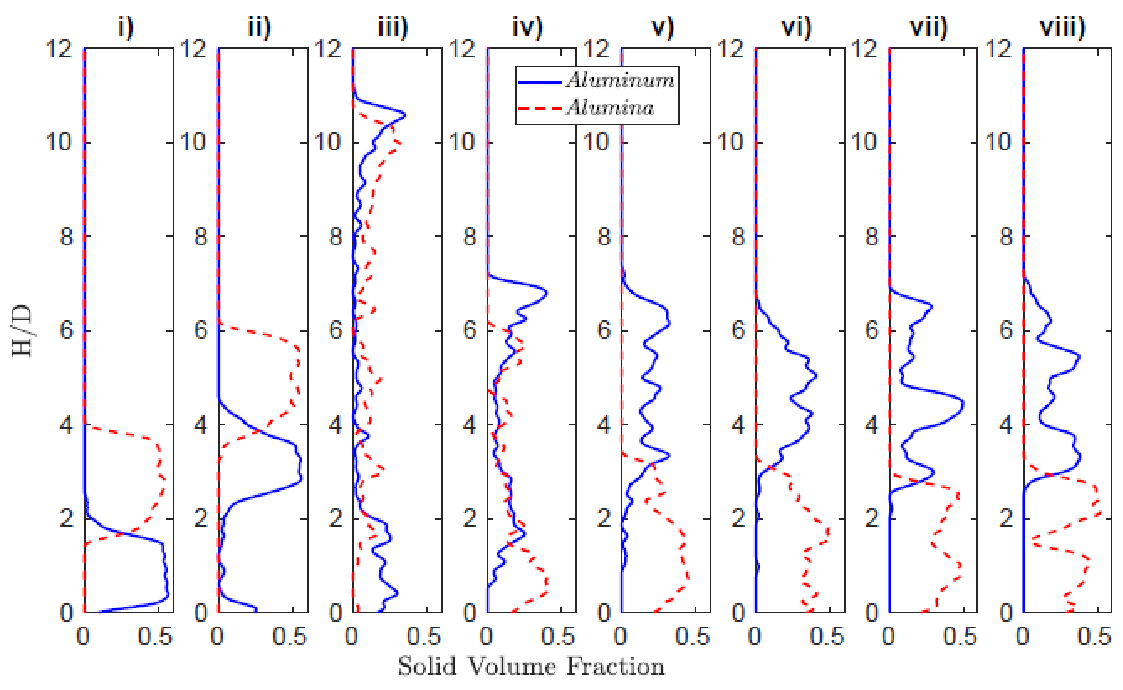}\\
 			(b)
 		\end{center}
 	\end{minipage}
	\hfill
	\caption{Solid volume fractions at different times for case II and (a) $\overline{U}$ = 0.137 m/s; (b) $\overline{U}$ = 0.164 m/s. The dashed red line corresponds to species 1 and the continuous blue line to species 2. The corresponding times are: (i) 0 s; (ii) 3 s; (iii) 6 s; (iv) 9 s; (v) 12 s; (vi) 15 s; (vii) 18 s; (viii) 21 s.}
	\label{fig:SVF_150_250}
 \end{figure}

We determined the volume fractions of each species by using numerical scripts \cite{Supplemental3} that compute the cross-section areas occupied by each grain type at given intervals of bed height, for all time steps. Fig. \ref{fig:SVF_150_250} presents profiles of mean volume fractions as a function of the bed height for case $II$, for both flow rates. The dashed red line corresponds to species 1 and the continuous blue line to species 2. The corresponding times are in the caption of figures. Cases $I$, $III$, and $IV$ present similar behaviors (the respective figures can be seen in the  Supplementary Material \cite{Supplemental}).

We observe from Fig. \ref{fig:SVF_150_250} that species start mixing as soon as the initial transient begins, occurring both within the initial plug and at the bottom of the bed.  By the end of the initial transient (snapshot (iv) in Fig. \ref{fig:SVF_150_250}a and (iii) in Fig. \ref{fig:SVF_150_250}b), a considerable degree of mixing is achieved, the volume fractions of both species having comparable values throughout the bed. After the initial transient, (snapshot (v) in Fig. \ref{fig:SVF_150_250}a and (iv) in Fig. \ref{fig:SVF_150_250}b), the mixing degree decreases, the volume fractions of species 1 and 2 becoming, respectively, higher at the bottom and on the top until the layers are completely inverted (snapshot (vii) in Fig. \ref{fig:SVF_150_250}a and (vi) in Fig. \ref{fig:SVF_150_250}b). In summary, mixing occurs mainly during the initial transient and segregation afterward.

The characteristic time for the inversion can be obtained from the profiles of volume fraction at different times. This time was computed for each simulated condition as the time that the particle volume fractions take to complete the inversion, based on data as plotted in Fig. \ref{fig:SVF_150_250}. We expect that the time for layer inversion varies with the bed height and inversely with the liquid velocity, the latter being responsible for the suspension of solids and the former representing an obstacle to surpass. From Fig. \ref{fig:SVF_150_250}, we observe that, indeed, the time for inversion varies with both $h_{mf}$ and the inverse of $\overline{U}$. We propose that $t_{scale} = h_{mf}/\overline{U}$ is the timescale for layer inversion; therefore, the characteristic time for layer inversion $t_{inv}$ is given by Eq. \ref{eq:time},

\begin{equation}
t_{inv} = A\, t_{scale} = A\, h_{mf}/\overline{U}
\label{eq:time}
\end{equation}

\noindent where $A$ is a dimensionless multiplicative factor. We note that the densities of the two species may also play a role in $t_{inv}$, their ratio $\rho_{p2}/\rho_{p1}$ affecting then $t_{inv}$. Because in this study we used only two fixed species, we preferred to not write explicitly $\rho_{p2}/\rho_{p1}$ in Eq. \ref{eq:time}, so that $\rho_{p2}/\rho_{p1}$ is embedded in $A$. From the volume fractions obtained numerically, we found the inversion time for each tested condition. The values of $A = t_{inv} / t_{scale}$ are shown in Tab. \ref{table:table_num}, from which we observe that $A \approx 20$. Therefore, the characteristic time for layer inversion, within the range of parameters covered by this study, is $t_{inv} = 20\, h_{mf}/\overline{U}$. To the authors' knowledge, the inversion time has never been quantified, so that we propose a new scaling for the duration of the inversion process.

\begin{figure}[ht]
 	\begin{minipage}{0.5\linewidth}
 		\begin{tabular}{c}
 			\includegraphics[width=0.90\linewidth]{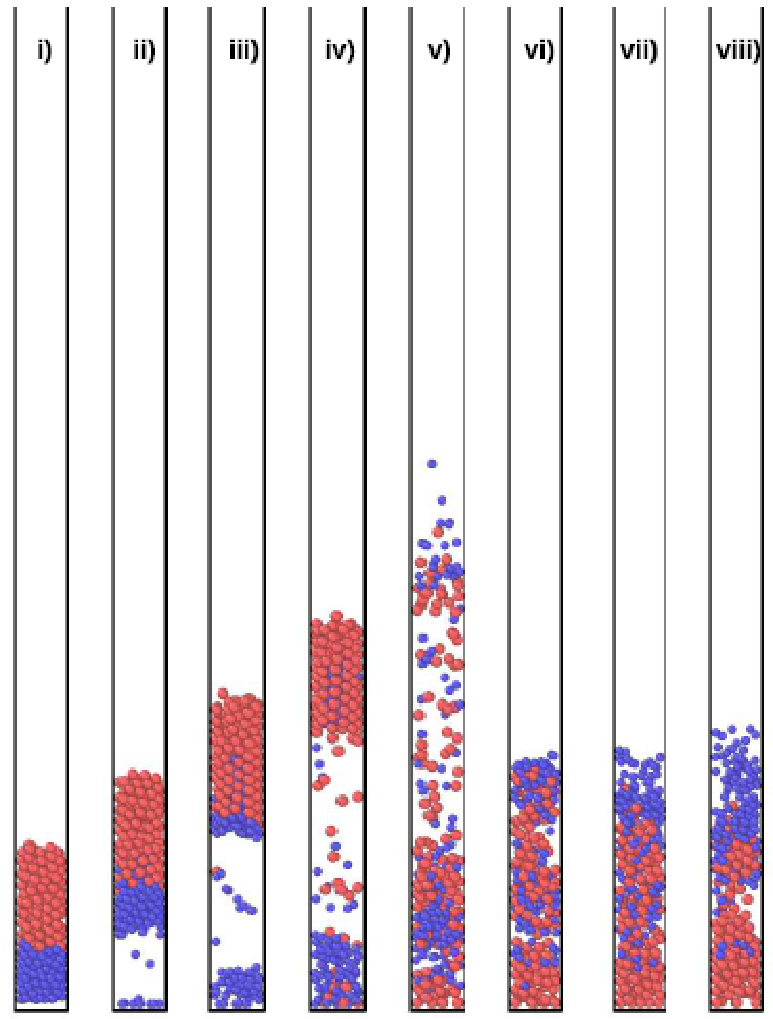}\\
 			(a)
 		\end{tabular}
 	\end{minipage}
 	\hfill
 	\begin{minipage}{0.5\linewidth}
 		\begin{tabular}{c}
 			\includegraphics[width=0.90\linewidth]{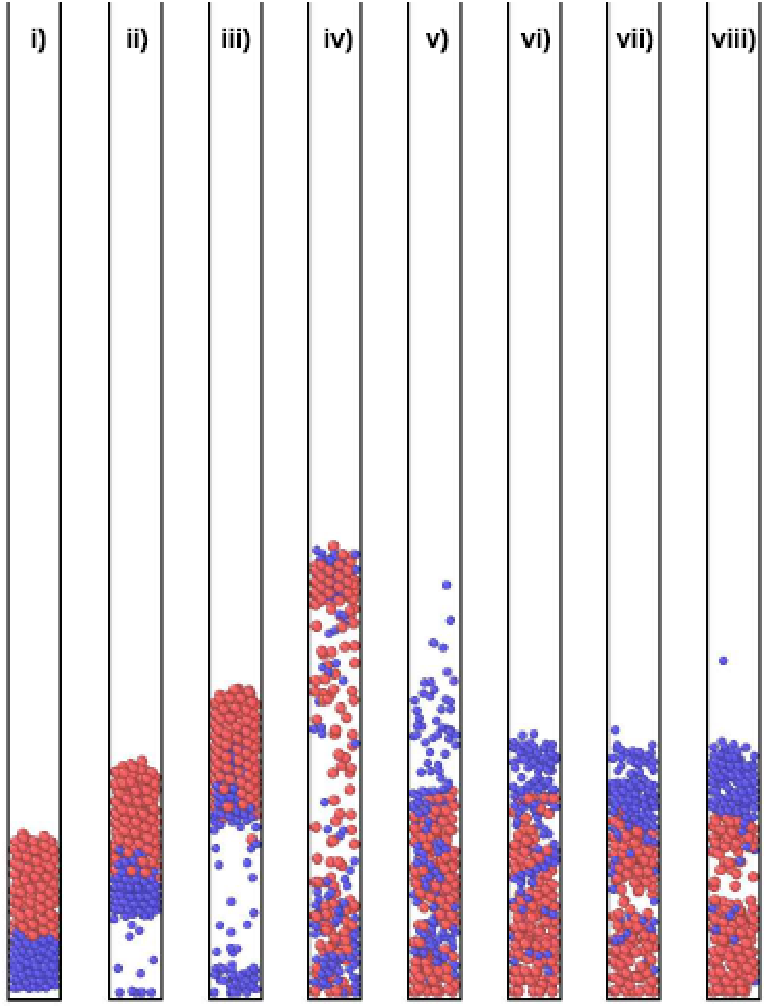}\\
 			(b)
 		\end{tabular}
 	\end{minipage}
	\hfill
	\caption{Instantaneous snapshots of particle positions for case I and $\overline{U}$ = 0.164 m/s. (a) With virtual mass force; (b) without virtual mass force. The corresponding times are: (i) 0 s; (ii) 1 s; (iii) 2 s; (iv) 3 s; (v) 4 s; (vi) 5 s; (vii) 6 s; (viii) 7 s.}
	\label{fig:snapshots_compara_VM}
 \end{figure}

\begin{figure}[ht]
 	\begin{minipage}[c]{\columnwidth}
 		\begin{center}
 			\includegraphics[width=0.90\linewidth]{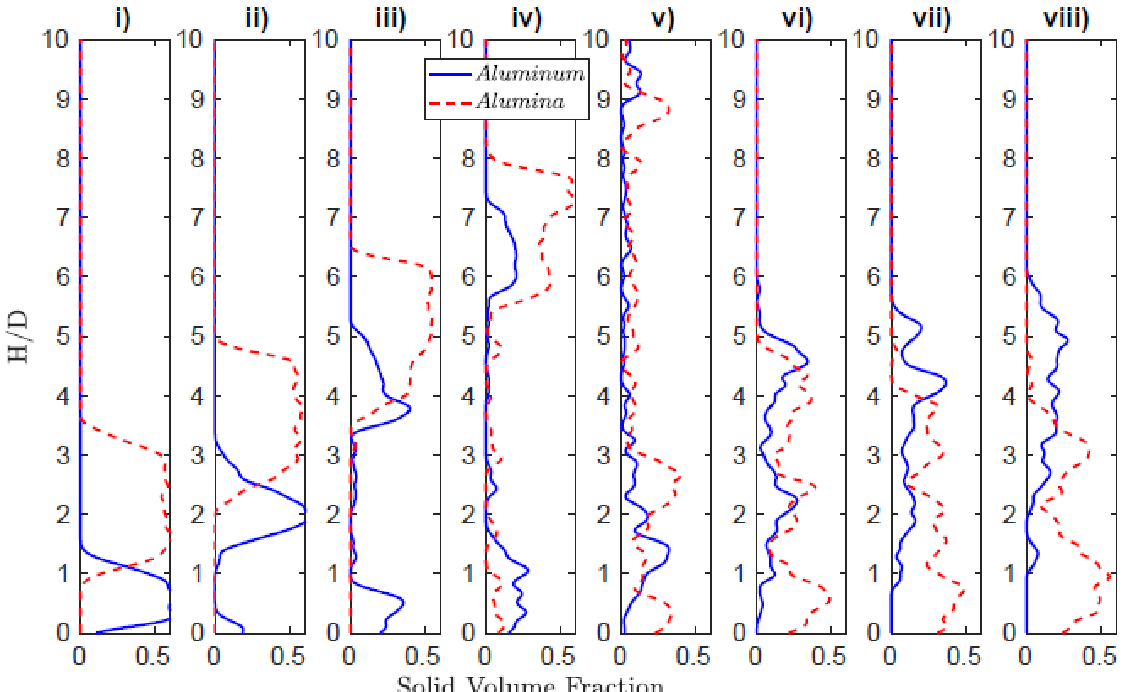}\\
 			(a)
 		\end{center}
 	\end{minipage}
 	\hfill
 	\begin{minipage}[c]{\columnwidth}
 		\begin{center}
 			\includegraphics[width=0.90\linewidth]{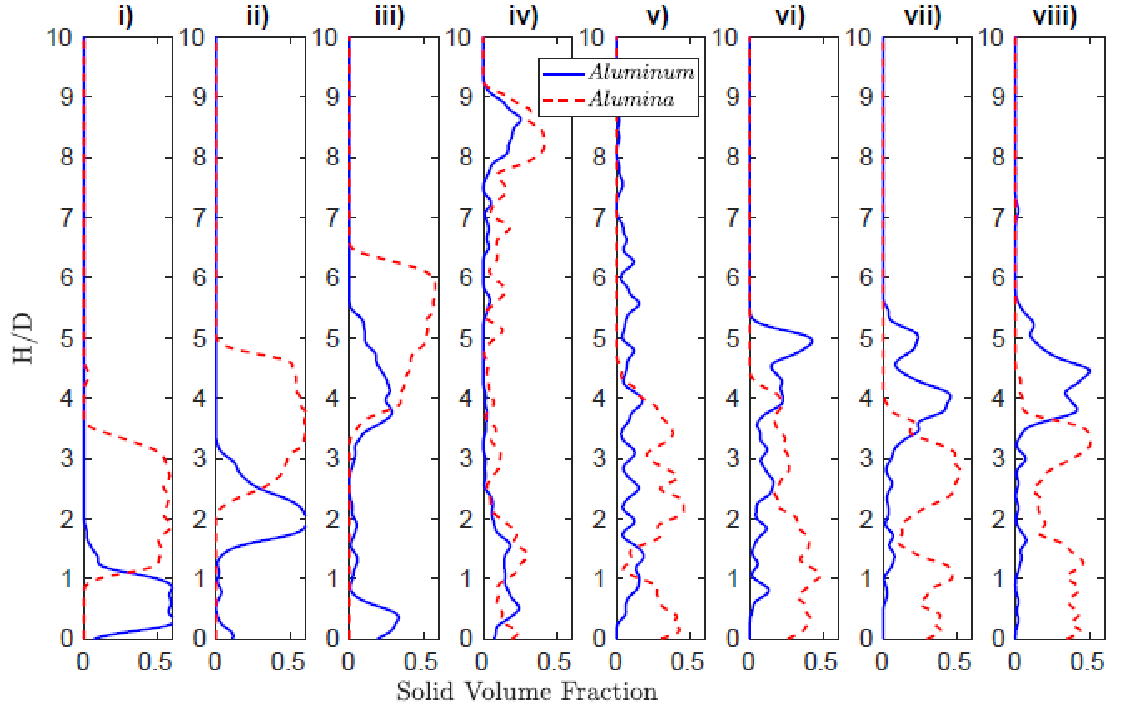}\\
 			(b)
 		\end{center}
 	\end{minipage}
	\hfill
	\caption{Solid volume fractions at different times for case I and $\overline{U}$ = 0.164 m/s. (a) With virtual mass force; (b) without virtual mass force. The dashed red line corresponds to species 1 and the continuous blue line to species 2. The corresponding times are: (i) 0 s; (ii) 1 s; (iii) 2 s; (iv) 3 s; (v) 4 s; (vi) 5 s; (vii) 6 s; (viii) 7 s.}
	\label{fig:SVF_compara_VM}
 \end{figure}

In order to investigate the importance of the virtual mass force in layer inversions in SLFB, we performed numerical simulations of case I with and without the virtual mass force. Figs. \ref{fig:snapshots_compara_VM}a and \ref{fig:snapshots_compara_VM}b present the instantaneous snapshots of particle positions for case I obtained from numerical simulations performed with and without the virtual mass force, respectively, and Figs. \ref{fig:SVF_compara_VM}a and \ref{fig:SVF_compara_VM}b present the solid volume fractions at different times for case I from numerical simulations with and without the virtual mass force, respectively. The corresponding fluid velocity and times are in the caption of figures. We note that by not considering the virtual mass force the time for layer inversion is smaller than that obtained from both experiments and simulations considering the virtual mass force. The virtual mass force corresponds to the acceleration of part of the fluid when the solid particle is accelerated through it. Therefore, this force is associated to an additional work done by the solid particle to accelerate part of the fluid, being of dissipative nature. When accounted for, this force slows down the inversion process. The present simulations capture this effect.

\begin{figure}
\centering
\begin{minipage}[c]{\textwidth}
\centering
\begin{tabular}{c}
\includegraphics[scale=0.50]{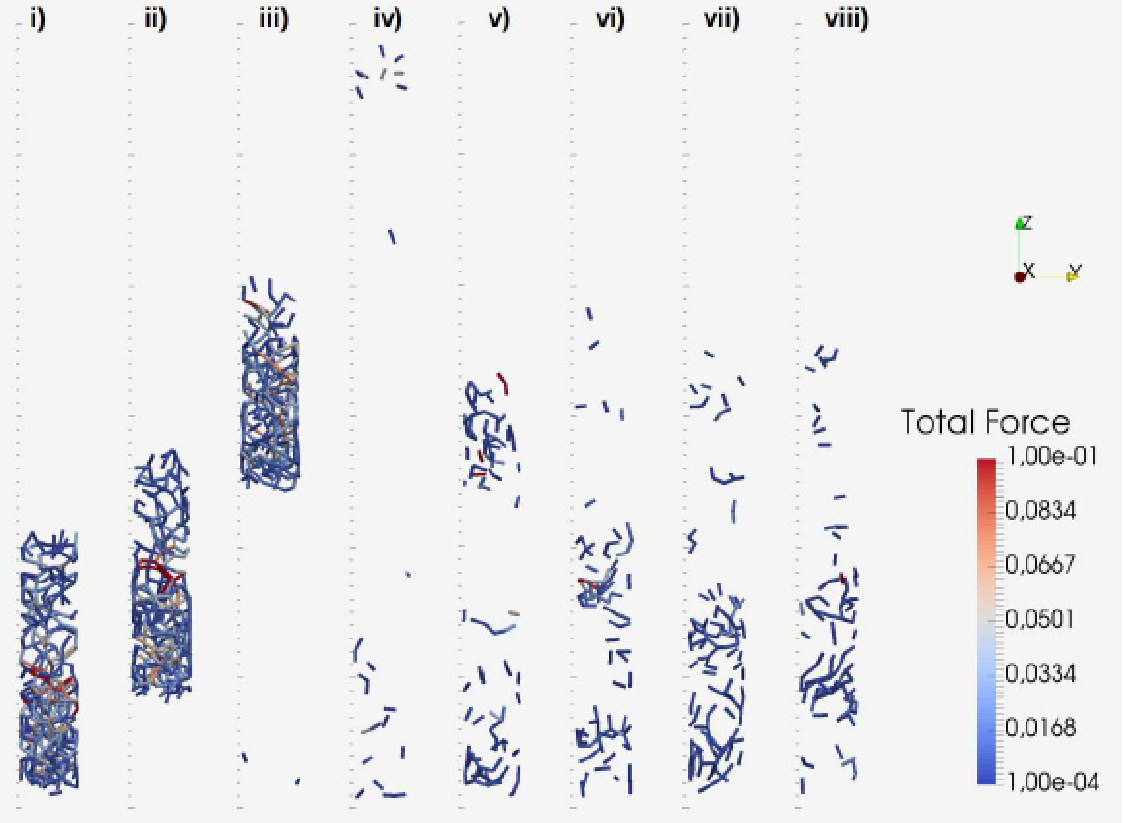}\\
      (a)
\end{tabular}
\end{minipage} \hfill
\begin{minipage}[c]{\textwidth}
\centering
\begin{tabular}{c}
\includegraphics[scale=0.50]{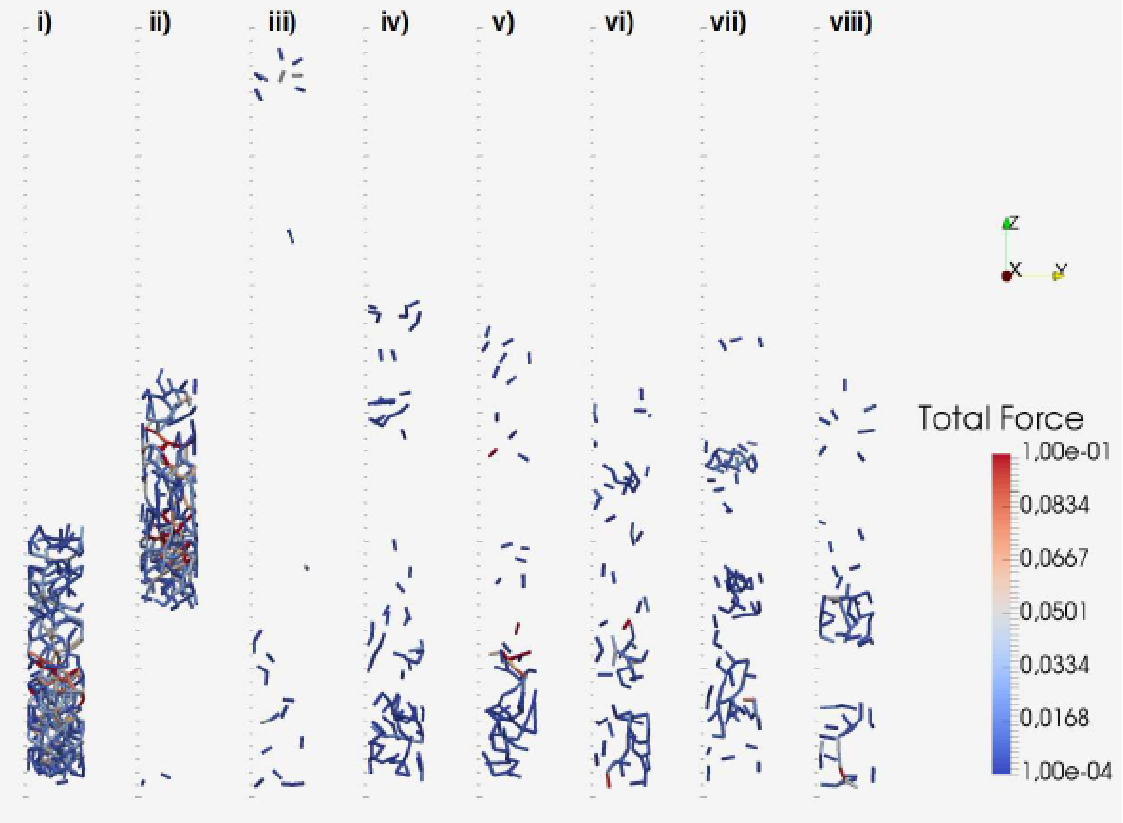}\\
      (b)
\end{tabular}
\end{minipage}
\caption{Instantaneous snapshots of the network of contact forces for case II and (a) $\overline{U}$ = 0.137 m/s.; (b) $\overline{U}$ = 0.164 m/s. The corresponding times are: (i) 0 s; (ii) 3 s; (iii) 6 s; (iv) 9 s; (v) 12 s; (vi) 15 s; (vii) 18 s; (viii) 21 s. Values are in N.}
\label{fig:snapshots_fc_caseII}	
\end{figure}

Finally, we present the evolution of the network of contact forces, which shows the influence of the lateral walls on layer inversion. The network of contact forces was obtained from the numerical simulations, being inaccessible from our experiments. Figs. \ref{fig:snapshots_fc_caseII}a and \ref{fig:snapshots_fc_caseII}b show instantaneous snapshots of the network of contact forces for case $II$ with $\overline{U}$ = 0.137 and 0.164 m/s, respectively, at the same instants depicted in Figs. \ref{fig:snapshots_150_250_250} and \ref{fig:snapshots_150_250_300}. Cases $I$, $III$, and $IV$ present similar behaviors (the respective figures can be seen in the Supplementary Material \cite{Supplemental}).  

Although we observe the presence of contact chains during all the inversion, they are much stronger at the initial transient, when the single granular plug is displaced upwards. This corroborates the observations made in Subsection \ref{section_experimental_results}, that the transient plug is caused by the high confinement characteristic of the very narrow case.

In summary, we showed the bed dynamics during the inversion of layers in narrow tubes by mimicking the layer inversion phenomenon, both experimentally and numerically. Our numerical simulations captured well the dynamics of individual particles during the inversion of layers in narrow tubes. The simulations were performed with the CFDEM, LIGGGHTS and OpenFOAM codes, where the liquid flow was computed with the two-phase flow equations and the motion of solids with linear and angular momentum equations. We found the distances traveled by grains during the inversion and the characteristic time for inversion. The numerical results are encouraging as the used method can be applied to more complex industrial scenarios.

\section{Conclusions}

This paper investigated experimentally and numerically the dynamics of solid particles during layer inversion in binary SLFB in narrow tubes. Confinement effects caused by the tube walls change the way in which individual particles move. Although layer inversion is present in industrial applications, experimental results on particle trajectories during the inversion were not reported in previous studies. In addition, no previous study, experimental or numerical, was conducted for the narrow case.

The fluidized beds were formed in a 25.4 mm-ID pipe and consisted of two different solid species: alumina beads with $d_1$ = 6 mm and $\rho_{p1}$ = 3690 kg/m$^3$ (species 1), and aluminum beads with $d_2$ = 4.8 mm and $\rho_{p2}$ = 2760 kg/m$^3$ (species 2). The ratios between the tube and grain diameters were of 4.23 and 5.29 for species 1 and 2, respectively, corresponding to a very narrow case. The species were chosen in order to force an inversion of layers, mimicking the layer inversion mechanism in the beginning of each run. For that, the initial beds were formed with the lighter particles on the bottom and the heavier ones on the top, for both the experiments and simulations. In our experiments, the fluidized beds were filmed with a high-speed camera, and individual particles were automatically identified and tracked along images by using numerical scripts. For the numerical part, we performed three-dimensional simulations using a coupled CFD-DEM code. The fluid flow was computed with the open source code OpenFOAM, while the granular dynamics was computed with the open source code LIGGGHTS, both linked via the open source code \mbox{CFDEM}. Different from previous numerical works on layer inversion, we considered the virtual mass force, which is important in the case of liquids. Our numerical results using CFD-DEM were able to capture the dynamics of individual particles during the inversion of layers in narrow tubes. 

We obtained the trajectories of particles, particle fractions and network of contact forces during the layer inversion in very narrow tubes from experiments and numerical simulations. Under the tested conditions, an initial transient occurs where a single plug rises while particles on its bottom fall freely. The ascending plug is the result of a blockage due to arches within the granular plug, as shown in terms of the network of contact forces. As the free falling particles reach the bottom of the tube and the initial plug decreases, some of the smaller particles move to the top of both the bottom bed and initial plug. After the initial plug has vanished and all grains reached the bed, the layer inversion continues.

During the inversion, particles of each species follow oscillating paths between the bottom and top regions of the bed. The vertical distances traveled by particles decrease with time, with the aluminum and alumina particles oscillating within the top and bottom regions, respectively, by the end of the inversion process. Within the range of parameters investigated, the typical distances traveled by particles during the layer inversion are of 5 to 6 $h_{mf}$ and 6.5 to 7.5 $h_{mf}$ for the alumina and aluminum beads, respectively.

Based on the numerical results, we computed the profiles of volume fraction of each species during the layer inversion. From them, we observed that mixing occurs mainly during the initial transient, while segregation occurs afterward. Species start mixing from the beginning of the initial transient, occurring both within the initial plug and at the bottom of the bed, and by the end of the initial transient a considerable degree of mixing is achieved. Segregation then takes place after the end of the initial transient until the layers are completely inverted. By considering the profiles of volume fraction, we found the typical time for complete inversion, which is $t_{inv} = 20\, h_{mf}/\overline{U}$ for all tested cases.

In conclusion, we obtained the complete behavior of individual grains during the layer inversion in narrow pipes from both experiments and numerical simulations. We found the distances traveled by particles during the inversion and the characteristic time for layer inversion. In addition, the numerical methodology was validated and can be applied to more complex industrial problems. The present findings represent a significant step toward understanding the layer inversion problem.

\subsection*{Nomenclature}
\begin{tabular}{l l}
$A$ & time coefficient (--)\\
$C_d$ & drag coefficient (--)\\
$D$ & tube diameter (m)\\
$d$ & grain diameter (m)\\
E & Young's modulus (Pa)\\
e & restitution coefficient (--)\\
$\vec{F}$ & force (N)\\
$\vec{F}_{c}$ & contact forces (N)\\
$\vec{F}_{fp}$ & drag force on grains (N)\\
$\vec{F}_{pf}$ & reaction on the fluid of the drag force on grains (N)\\
$h$ & bed height (m)\\
$H$ & height (m)\\
$\vec{g}$ & acceleration of gravity (m s$^{-2}$)\\
$I$ & moment of inertia (kg m$^2$)\\
$k$ & stiffness coefficient (N m$^{-1}$)\\
$m$ & mass (kg)\\
$N$ & number of particles (--)\\
$\vec{n}$ & unit vector in the normal direction (--)\\
$P$ & pressure (Pa)\\
$Q$ & volumetric flow rate of the fluid (m$^3$ s$^{-1}$)\\
$Re_{D}$ & Reynolds number based on the superficial velocity and tube diameter (--)\\
$Re_{d}$ & Reynolds number based on the superficial velocity and grain diameter (--)\\
$Re_{t}$ & Reynolds number based on the terminal velocity and grain diameter (--)\\
$S$ & solid to fluid density ratio (--)\\
$S_t$ & Stokes number based on the terminal velocity (--)\\

\end{tabular}

\begin{tabular}{l l}
$\vec{T}$ & torque (N m)\\
$t$ & time (s)\\
$\vec{t}$ & unit vector in the tangential direction (--)\\
$\overline{U}$ & superficial velocity of the fluid (m s$^{-1}$)\\
$U_f$ & fluid velocity through the packed bed (m s$^{-1}$)\\
$\vec{u}$ & velocity vector (m s$^{-1}$)\\
$\vec{u}_f$ & fluid velocity (m s$^{-1}$)\\
$\vec{u}_{fp}$ & fluid velocity at the grain position (m s$^{-1}$)\\
$\vec{u}_p$ & velocity of each solid particle (m s$^{-1}$)\\
$\vec{u}_{sij}$ & slip velocity at the contact between particles i and j (m s$^{-1}$)\\
$V_{cell}$ & volume of a computational cell (m$^3$)\\
$V_{p}$ & volume of a solid particle (m$^3$)\\
$v_t$ & terminal velocity (m s$^{-1}$)\\
$v_s$ & settling velocity (m s$^{-1}$)\\
$x$ & radial coordinate (m)\\
$z$ & vertical coordinate (m)\\
\end{tabular}

\subsection*{Greek symbols}
\begin{tabular}{l l}
$\beta$ & coefficient of drag force (--)\\
$\delta$ & deformation distance (m)\\
$\eta$ & damping coefficient (kg s$^{-1}$)\\
$\mu$ & dynamic viscosity (Pa s)\\
$\mu_{fr}$ & friction coefficient (--)\\
$\rho$ & density (kg m$^{-3}$)\\
$\sigma$ & Poisson ratio (--)\\
$\vec{\vec{\tau}}$ & shear stress tensor (N m$^{-2}$)\\
\end{tabular}

\subsection*{Subscripts}
\begin{tabular}{l l}
1 & relative to species 1\\
2 & relative to species 2\\
$c$ & relative to contacts\\
$d$ & relative to the grain diameter\\
$D$ & relative to the tube diameter\\
$f$ & relative to the fluid\\
$i$ & particle index\\
$int$ & relative to the interface between species 1 and 2\\
$inv$ & relative to the inversion\\
$j$ & particle index\\
$mf$ & relative to minimum fluidization\\
$n$ & normal\\
$p$ & relative to solid particles\\
$press$ & relative to fluid stresses\\
$s$ & relative to settling velocity\\
$scale$ & relative to a proposed scale\\
$t$ & relative to terminal velocity or also tangential\\
$top$ & relative to the top of the bed\\
$vm$ & relative to virtual mass\\
$w$ & relative to the tube wall\\
\end{tabular}

\subsection*{Superscripts}
\begin{tabular}{l l}
$\overline{\quad}$ & cross-section average\\
\end{tabular}

\section*{Acknowledgments}
 
Fernando David C\'u\~nez Benalc\'azar is grateful to FAPESP (Grant No. 2016/18189-0), and Erick de Moraes Franklin would like to express his gratitude to FAPESP (Grants No. 2016/13474-9 and No. 2018/14981-7), to CNPq (grant no. 400284/2016-2) and to FAEPEX/UNICAMP (Grant No. 2112/19) for the financial support they provided.

%% file: main.bbl
\begin{thebibliography}{10}
\expandafter\ifx\csname url\endcsname\relax
  \def\url#1{\texttt{#1}}\fi
\expandafter\ifx\csname urlprefix\endcsname\relax\def\urlprefix{URL }\fi
\expandafter\ifx\csname href\endcsname\relax
  \def\href#1#2{#2} \def\path#1{#1}\fi

\bibitem{Epstein_1}
N.~Epstein, B.~P. Leclair, B.~Pruden, Liquid fluidization of binary particle
  mixtures -- {I}: {O}verall bed expansion", Chem. Eng. Sci. 36~(11) (1981)
  1803 -- 1809.

\bibitem{Epstein_3}
N.~Epstein, Teetering, Powder Technol. 151~(1) (2005) 2 -- 14.

\bibitem{Hancock}
R.~Hancock, The teeter condition, Mining Magazine 55 (1936) 90--94.

\bibitem{Epstein_2}
N.~Epstein, B.~P. Leclair, Liquid fluidization of binary particle mixtures --
  {II}: {B}ed inversion", Chem. Eng. Sci. 40~(8) (1985) 1517 -- 1526.

\bibitem{Gibilaro}
L.~Gibilaro, R.~D. Felice, S.~Waldram, P.~Foscolo, A predictive model for the
  equilibrium composition and inversion of binary-solid liquid fluidized beds,
  Chem. Eng. Sci. 41~(2) (1986) 379 -- 387.

\bibitem{Moritomi}
H.~Moritomi, T.~Iwase, T.~Chiba, A comprehensive interpretation of solid layer
  inversion in liquid fluidised beds, Chem. Eng. Sci. 37~(12) (1982) 1751 --
  1757.

\bibitem{Chun}
B.-S. Chun, D.~H. Lee, N.~Epstein, J.~R. Grace, A.-H.~A. Park, S.~D. Kim, J.~K.
  Lee, Layer inversion and mixing of binary solids in two- and three-phase
  fluidized beds, Chem. Eng. Sci. 66~(14) (2011) 3180 -- 3184.

\bibitem{Dimaio}
F.~P. {Di Maio}, A.~{Di Renzo}, Direct modeling of voidage at layer inversion
  in binary liquid-fluidized bed, Chem. Eng. J. 284 (2016) 668 -- 678.

\bibitem{Cornelissen}
J.~T. Cornelissen, F.~Taghipour, R.~Escudié, N.~Ellis, J.~R. Grace, {CFD}
  modelling of a liquid-solid fluidized bed, Chem. Eng. Sci. 62~(22) (2007)
  6334 -- 6348.

\bibitem{wang}
S.~Wang, X.~Li, Y.~Wu, X.~Li, Q.~Dong, C.~Yao, Simulation of flow behavior of
  particles in a liquid-solid fluidized bed, Ind. Eng. Chem. Res. 49 (2010)
  10116--10124.

\bibitem{Patankar}
N.~A. Patankar, D.~D. Joseph, Modeling and numerical simulation of particulate
  flows by the {Eulerian-Lagrangian} approach, Int. J. Multiphase Flow 27~(10)
  (2001) 1659 -- 1684.

\bibitem{Ghatage}
S.~V. Ghatage, Z.~Peng, M.~J. Sathe, E.~Doroodchi, N.~Padhiyar, B.~Moghtaderi,
  J.~B. Joshi, G.~M. Evans, Stability analysis in solid-liquid fluidized beds:
  Experimental and computational, Chem. Eng. J. 256 (2014) 169 -- 186.

\bibitem{Alobaid}
F.~Alobaid, An offset-method for {Euler-Lagrange} approach, Chem. Eng. Sci. 138
  (2015) 173 -- 193.

\bibitem{Chiesa}
M.~Chiesa, V.~Mathiesen, J.~A. Melheim, B.~Halvorsen, Numerical simulation of
  particulate flow by the {Eulerian-Lagrangian} and the {Eulerian-Eulerian}
  approach with application to a fluidized bed, Comput. Chem. Eng. 29~(2)
  (2005) 291 -- 304.

\bibitem{Vegendla}
S.~N.~P. Vegendla, G.~J. Heynderickx, G.~B. Marin, Comparison of
  {Eulerian-Lagrangian} and {Eulerian-Eulerian} method for dilute gas–solid
  flow with side inlet, Comput. Chem. Eng. 35~(7) (2011) 1192 -- 1199.

\bibitem{Adamczyk}
W.~P. Adamczyk, A.~Klimanek, R.~A. Białecki, G.~Węcel, P.~Kozołub,
  T.~Czakiert, Comparison of the standard {Euler-Euler} and hybrid
  {Euler-Lagrange} approaches for modeling particle transport in a pilot-scale
  circulating fluidized bed, Particuology 15 (2014) 129 -- 137.

\bibitem{cundall1979discrete}
P.~A. Cundall, O.~D. Strack, A discrete numerical model for granular
  assemblies, G\'eotechnique 29~(1) (1979) 47--65.

\bibitem{Kloss}
C.~Kloss, C.~Goniva, {LIGGGHTS}: a new open source discrete element simulation
  software, in: Proc. 5th Int. Conf. on Discrete Element Methods, London, UK,
  2010.

\bibitem{Berger}
R.~Berger, C.~Kloss, A.~Kohlmeyer, S.~Pirker, Hybrid parallelization of the
  {LIGGGHTS} open-source {DEM} code, Powder Technol. 278 (2015) 234--247.

\bibitem{Li}
T.~Li, H.~Zhang, M.~Liu, Z.~Huang, H.~Bo, Y.~Dong, {DEM} study of granular
  discharge rate through a vertical pipe with a bend outlet in small absorber
  sphere system, Nucl. Eng. Des. 314 (2017) 1--10.

\bibitem{Sun}
X.~Sun, M.~Sakai, Three-dimensional simulation of gas-solid-liquid flows using
  the {DEM-VOF} method, Chem. Eng. Sci. 134 (2015) 531--548.

\bibitem{Sun2}
R.~Sun, H.~Xiao, Sedi{F}oam: A general-purpose, open-source {CFD-DEM} solver
  for particle-laden flow with emphasis on sediment transport, Comput. Geosci.
  89 (2016) 207--219.

\bibitem{Liu}
G.~Liu, F.~Yu, H.~Lu, S.~Wang, P.~Liao, Z.~Hao, {CFD-DEM} simulation of
  liquid-solid fluidized bed with dynamic restitution coefficient, Powder
  Technol. 304 (2016) 186--197.

\bibitem{li2017modeling}
L.~Li, B.~Li, Z.~Liu, Modeling of spout-fluidized beds and investigation of
  drag closures using openfoam, Powder Technol. 305 (2017) 364--376.

\bibitem{Reddy}
R.~K. Reddy, J.~B. Joshi, {CFD} modeling of solid-liquid fluidized beds of mono
  and binary particle mixtures, Chem. Eng. Sci. 64~(16) (2009) 3641 -- 3658.

\bibitem{Peng}
Z.~Peng, J.~B. Joshi, B.~Moghtaderi, M.~S. Khan, G.~M. Evans, E.~Doroodchi,
  Segregation and dispersion of binary solids in liquid fluidised beds: {A}
  {CFD-DEM} study, Chem. Eng. Sci. 152 (2016) 65 -- 83.

\bibitem{Galvin}
K.~P. Galvin, R.~Swann, W.~F. Ramirez, Segragation and dispersion of a binary
  system of particles in a fluidized bed, Chem. Eng. J. 52~(10) (2006)
  3401--3410.

\bibitem{Khan}
M.~S. Khan, S.~Mitra, S.~V. Ghatage, E.~Doroodchi, J.~B. Joshi, G.~M. Evans,
  Segregation and dispersion studies in binary solid-liquid fluidised beds: {A}
  theoretical and computational study, Powder Technol. 314 (2017) 400 -- 411.

\bibitem{Abbaszadeh}
E.~{Abbaszadeh Molaei}, A.~Yu, Z.~Zhou, Investigation of causes of layer
  inversion and prediction of inversion velocity in liquid fluidizations of
  binary particle mixtures, Powder Technol. 342 (2019) 418 -- 432.

\bibitem{Nelson}
M.~J. Nelson, G.~Nakhla, J.~Zhu, Fluidized-bed bioreactor applications for
  biological wastewater treatment: {A} review of research and developments,
  Engineering 3~(3) (2017) 330 -- 342.

\bibitem{Zenit}
R.~Zenit, M.~L. Hunt, C.~E. Brennen, Collisional particle pressure measurements
  in solid-liquid flows, J. Fluid Mech. 353 (1997) 261--283.

\bibitem{Zenit2}
R.~Zenit, M.~L. Hunt, Solid fraction fluctuations in solid-liquid flows, Int.
  J. Multiphase Flow 26~(5) (2000) 763 -- 781.

\bibitem{Aguilar}
A.~Aguilar-Corona, R.~Zenit, O.~Masbernat, Collisions in a liquid fluidized
  bed, Int. J. Multiphase Flow 37~(7) (2011) 695 -- 705.

\bibitem{Cunez}
F.~D. C\'u\~nez, E.~M. Franklin, Plug regime in water fluidized beds in very
  narrow tubes, Powder Technol. 345 (2019) 234--246.

\bibitem{Goniva}
C.~Goniva, C.~Kloss, N.~G. Deen, J.~A.~M. Kuipers, S.~Pirker, Influence of
  rolling friction on single spout fluidized bed simulation, Particuology
  10~(5) (2012) 582--591.

\bibitem{gidaspow1994multiphase}
D.~Gidaspow, Multiphase flow and fluidization: continuum and kinetic theory
  descriptions, Academic press, 1994.

\bibitem{Zhou}
Z.~Y. Zhou, S.~B. Kuang, K.~W. Chu, A.~B. Yu, Discrete particle simulation of
  particle–fluid flow: model formulations and their applicability, J. Fluid
  Mech. 661 (2010) 482–510.

\bibitem{tsuji1992lagrangian}
Y.~Tsuji, T.~Tanaka, T.~Ishida, Lagrangian numerical simulation of plug flow of
  cohesionless particles in a horizontal pipe, Powder Technol. 71~(3) (1992)
  239--250.

\bibitem{Kloss2}
C.~Kloss, C.~Goniva, A.~Hager, S.~Amberger, S.~Pirker, Models, algorithms and
  validation for opensource {DEM} and {CFD-DEM}, Prog. Comput. Fluid Dyn. An
  Int. J. 12~(2/3) (2012) 140--152.

\bibitem{Mondal}
S.~Mondal, C.-H. Wu, M.~M. Sharma, Coupled {CFD-DEM} simulation of hydrodynamic
  bridging at constrictions, Int. J. Multiphase Flow 84 (2016) 245 -- 263.

\bibitem{Davim}
J.~P. Davim, E.~Santos, C.~Pereira, J.~Ferreira, Comparative study of friction
  behaviour of alumina and zirconia ceramics against steel under water
  lubricated conditions, Ind. Lubr. Tribol. 60~(4) (2008) 178--182.

\bibitem{Supplemental}
See Supplementary Material for instantaneous snapshots of particle positions,
  particle fractions and network of contact forces, and a movie showing the
  evolution of a fluidized bed, both from experiments and numerical
  simulations. The movie corresponds to a bed consisting of 150 beads of each
  species fluidized by a water flow with superficial velocity of 0.164 m/s. The
  LHS corresponds to the images from the experiments and the RHS to the images
  from numerical simulations. The reproduction rate is half the real time.

\bibitem{Supplemental3}
F.~D. C\'u\~nez, E.~M. Franklin, {F}iles of numerical results, movies of all
  test runs, numerical scripts for image processing and numerical scripts for
  post-processing of numerical results are available at Mendeley Data,
  http://dx.doi.org/10.17632/4wss2f25yt.1 (2018).

\end{thebibliography}
